\documentclass{aa}  

\usepackage{graphicx}
\usepackage{txfonts}
\usepackage{xspace}
\usepackage{epsfig}
\usepackage{epstopdf}
\usepackage{natbib}
\usepackage{subfigure}
\usepackage{color}

\newcommand{\modif}[1]{#1}
\newcommand{\modiff}[1]{#1}

\newcommand{\uv}{\ensuremath{\{u, \mathrm{v}\}}-plane\xspace}
\newcommand{\uvs}{\ensuremath{\{u, \mathrm{v}\}}-planes\xspace}
\newcommand{\VV}{\ensuremath{V^2}\xspace}

\newcommand{\fso}{\ensuremath{f_0^*}\xspace}
\newcommand{\denv}{\ensuremath{d_\mathrm{env}}\xspace}
\begin{document}

   \title{A disk asymmetry in motion around the B[e] star MWC158\thanks{Based on observations performed with PIONIER mounted on the ESO \textit{Very Large Telescope interferometer} (programmes: 089.C-0211, 190.C-0963).}}


    \author{J. Kluska 
          \inst{1,2}
          \and
          M. Benisty\inst{2}
          \and
          F. Soulez\inst{3} 
          \and
          J.-P. Berger\inst{2,4}
          \and
          J.-B. Le Bouquin\inst{2}
          \and
          F. Malbet \inst{2}             
         \and
         B. Lazareff\inst{2}         
             \and
          E. Thi\'ebaut\inst{5}
          }

   \institute{University of Exeter, School of Physics, Stocker Road, Exeter, EX4 4QL, UK.\\
              \email{jkluska@astro.ex.ac.uk}
         \and
         Univ. Grenoble Alpes, IPAG, F-38000 Grenoble, France; CNRS, IPAG, F-38000 Grenoble, France.\\
                     \and        
             Biomedical imaging Group, Ecole polytechnique f\'ed\'erale de Lausanne, Switzerland \\
           \and
           European Southern Observatory, 85748, Garching by Munchen, Germany\\
              \and
             Univ. Lyon 1, Observatoire de Lyon, 9 avenue Charles Andre, F-69230 Saint-Genis Laval, France; CNRS, CRAL, F-69230 Saint-
Genis Laval, France; Ecole Normale Superieure de Lyon, F-69007 Lyon, France.
             }

   \date{}

 
  \abstract
   {MWC158 is a star with the B[e] phenomenon that shows strong spectrophotometric variability (in lines and in UV and visible continuum) attributed to phases of shell ejection. The evolutionary stage of this star was never clearly determined. Previous interferometric, spectropolarimetric and spectro-interferometric studies suggest a disk morphology for its environment.}
   {We investigate the origin of the variability within the inner astronomical unit of the central star using near-infrared interferometric observations with PIONIER at the VLTI over a two-year period. }
   {We performed an image reconstruction of the circumstellar environment using the SPARCO method. We discovered that the morphology of the circumstellar environment could vary on timescales of weeks or days. We carried out a parametric fit of the data with a model consisting of a star, a disk and a bright spot that represents a brighter emission in the disk.}
   {We detect strong morphological changes in the first astronomical unit around the star, that happen on a timescale of few months. We cannot account for such variability well with a binary model. Our parametric model fits the data well and allows us to extract the location of the asymmetry for different epochs.}
   {For the first time, we detect a morphological variability in the environment of MWC158. This variability is reproduced by a model of a disk and a bright spot. The locations of the bright spot suggest that it is located in the disk, but its precise motion is not determined. The origin of the asymmetry in the disk is complex and may be related to asymmetric shell ejections.}


   \maketitle
%

\section{Introduction}

Stars showing the B[e] phenomenon are defined by the presence of permitted (Balmer and FeII) and forbidden emission lines along with a large infrared excess indicating circumstellar (CS) dusty material. 
\citet{Lamers1998} classified these stars in five categories including supergiants, pre-main-sequence stars, planetary nebulae, symbiotic stars, and unclassified stars.
The diversity in the nature of stars showing the B[e] phenomenon is puzzling.
It is still not clear if this phenomenon is caused by the same physical process because only a few objects have been studied in depth.
Optical interferometry is a powerful tool to investigate the close environment of these stars \citep{Weigelt2011,Wheelwright2012,Wheelwright2013,Vural2014} and in some cases reveal their binarity \citep[e.g. ][]{Millour2009,Kraus2012A,Wang2012}.
Such a secondary component can disturb the environment of the primary star causing forbidden lines to appear.
The most studied B[e] stars, however, remain unclassified because of their unknown and strongly debated evolutionary status. 
Some of these stars show hints of a pre-main-sequence nature but are not associated with any star-forming region.
Studying the close environment of these massive and puzzling B[e] stars may bring new insights into our understanding of the baffling phenomena occurring around them.

\begin{table}[t]
 \caption{Stellar parameters of MWC158. }
   \centering
   \begin{tabular}{@{} c|rr@{}} 
   Param. & Value & Ref. \\
   \hline
        RA (J2000) & 06 51 33.40 & (1)\\ 
        DEC (J2000) & -06 57 59.45 & (1)\\
        Mass & 6 M$_\odot$  & (2), (3)\\
        Sp. Type & B7III &(2), (3)\\
      $d$ & 392$^{+110}_{-71}$\,pc &  (1) \\
      V & 6.6 & (2), (4), (5) \\
      H & 5.1 & (6)
   \end{tabular}
   \label{tab:mwc158}
   \tablebib{(1) \citet{vanLeeuwen_2007}, (2) \citet{BorgesI}, (3) \citet{Ellerbroek2015}, (4) \citet{Harrington_2009}, (5) \citet{Tomasella_2010}, (6) \citet{Cutri2003}.}
\end{table}

MWC158 (also called HD50138 and V743 Mon) is one of the brightest B[e] stars (see Table.\,\ref{tab:mwc158}) 
even though  it remains unclassified \citep[e.g.][hereafter E15]{Lamers1998, Ellerbroek2015}.
It was previously argued to be a young object \citep{Morrison1995} and an evolved object \modif{in the main sequence or close to its ending} \citep{BorgesI}. 
\modif{Its} lack of membership to a star-forming region \citep{Maddalena1986} is the main obstacle to the classification of this object as a pre-main-sequence star.

MWC158 shows strong line profile variability \citep[from years to hours,][]{Hutsemekers1985,Andrillat1991}, which has been linked to shell ejections or formation of a one-arm, ejection spiral in its environment \citep{BorgesI} that could indicate an evolved nature.
\citet{Pogodin1997} has reported variability of H$_\alpha$, helium and sodium lines on timescales from days to months.
\citet{Hutsemekers1985} has reported variability in the UV that is associated with an outburst or a shell phase \citep{Andrillat1991}.
Recently, variability was reported in the visible spectra with timescales ranging from days \citep[interpreted as stellar pulsations by][]{BorgesIII} to months \citep[linked with its environment, e.g.][E15]{BorgesI}.
The discussion on the variability of this object \modif{has been}  \modif{based on} spectrophotometry and \modif{until now there have been} no spatially resolved constraints.

The disk-like morphology of the CS environment was first indicated by polarimetric and spectro-polarimetric observations \citep{Vaidya1994,Bjorkman1998,Oudmaijer1999,Harrington2007} before they were confirmed by interferometric observations.
These observations were performed in the near- and mid-infrared focusing on the size and orientation of the object \citep{Monnier_2009,BorgesII}. 
The inclination \modif{i}s found to be around 50\degr-55\degr degrees and the position angle (P.A.) around 70\degr. 
These studies confirmed the presence of a disk morphology but could not explore its shape in detail.

Here we report the analysis of a new and rich dataset taken with PIONIER/VLTI and spanning different epochs. It allows us to image the inner parts of the disk in more detail and to look for morphological variability.
We present our observations in Sect.\,\ref{sec:obs}. 
In Sect.\,\ref{sec:imgrec} we describe the image reconstructions on the interferometric data.
In Sect.\,\ref{sec:binfit} we describe the parametric modelling of the interferometric data to take into account the observed morphological variability.
We discuss our results in Sect.\,\ref{sec:dis} and we conclude in Sect.\,\ref{sec:ccl}.


\section{Observations}
\label{sec:obs}

We performed PIONIER/VLTI observations of MWC158 to reach high angular resolution in the near-infrared continuum. 
The PIONIER instrument \citep{LeBouquin2011} combines four telescopes in the $H$-band and provides six \modif{squared} visibility \modif{(\VV)} measurements and four closure phases \modif{(CP)}. 
If the object is resolved well, the AT relocation capability can be used to pave  the \uv efficiently and enable image reconstruction.
The preparation of observations was carried out with \texttt{ASPRO2} \citep{Bourges2013}.
The calibrators were found using \texttt{SearchCal} \citep[][HD46931, HD47559, HD50639, HD51914, HD52265]{Bonneau2011}.
The data was reduced with \texttt{pndrs} \citep{LeBouquin2011}.

The log of observations is shown in Appendix \ref{app:log2} (Table~\ref{tab:log2}). 
These observations were taken during the PIONIER Science Verification programme, the PIONIER Herbig Ae/Be Large Programme (190.C-0963, PI: Berger) and during 089.C-0211 (PI: Benisty).
\begin{figure}[!t] 
   \centering
   \includegraphics[width=6cm]{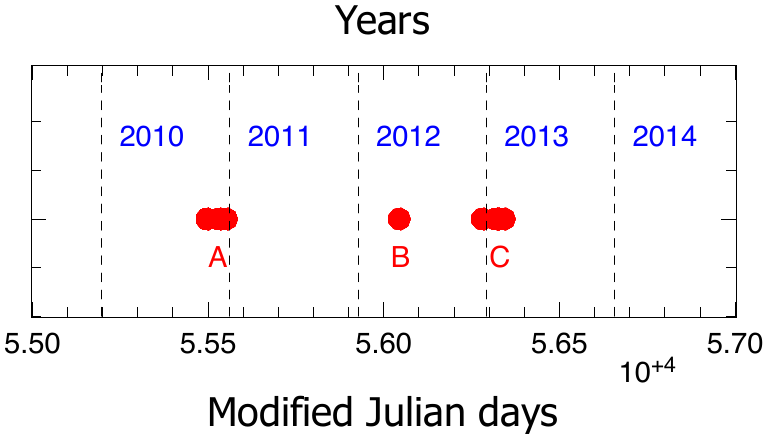} 
   \caption{Temporal representation on data taken on MWC158 during 3 epochs.}
   \label{fig:data_temp}
\end{figure}
The temporal distribution of all the observations is show\modif{n} \modif{in} Fig.~\ref{fig:data_temp}.
The target was observed in October and December 2010 (epoch A)\modif{,} April (epoch B) and December 2012, and January and February 2013 (epoch C).

\begin{figure}[!t] 
   \centering
   \includegraphics[width=5cm]{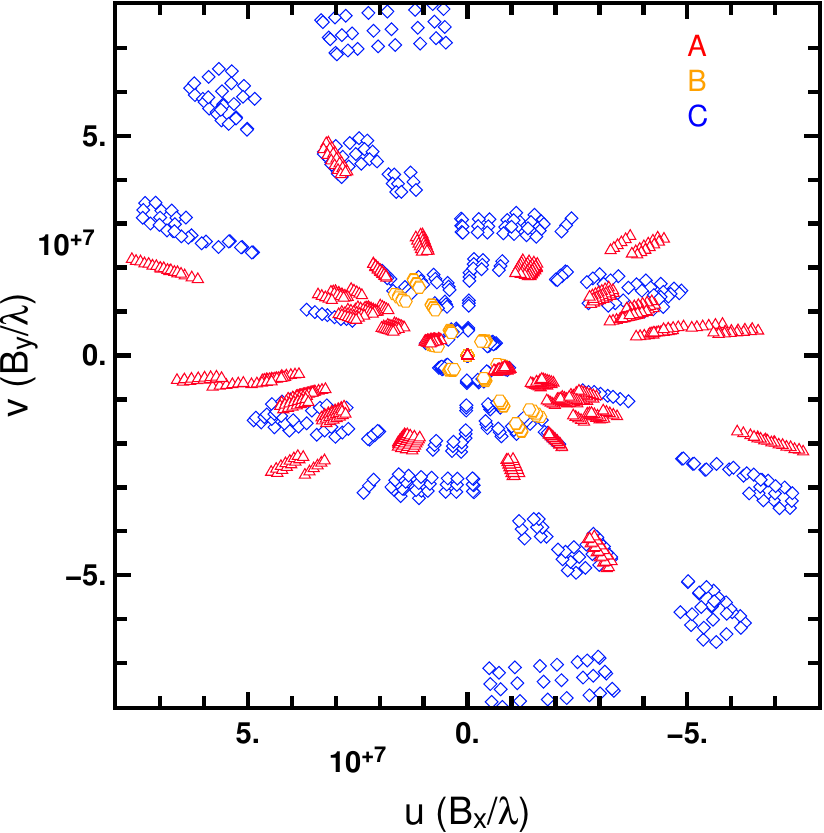} 
   \caption{ \uv for the 3 epochs. Epoch A is indicated in red triangles, epoch B in orange circles, and epoch C in blue diamonds.}
   \label{fig:uv}
\end{figure}

\begin{figure}[t] 
   \centering
   \includegraphics[width=9.2cm]{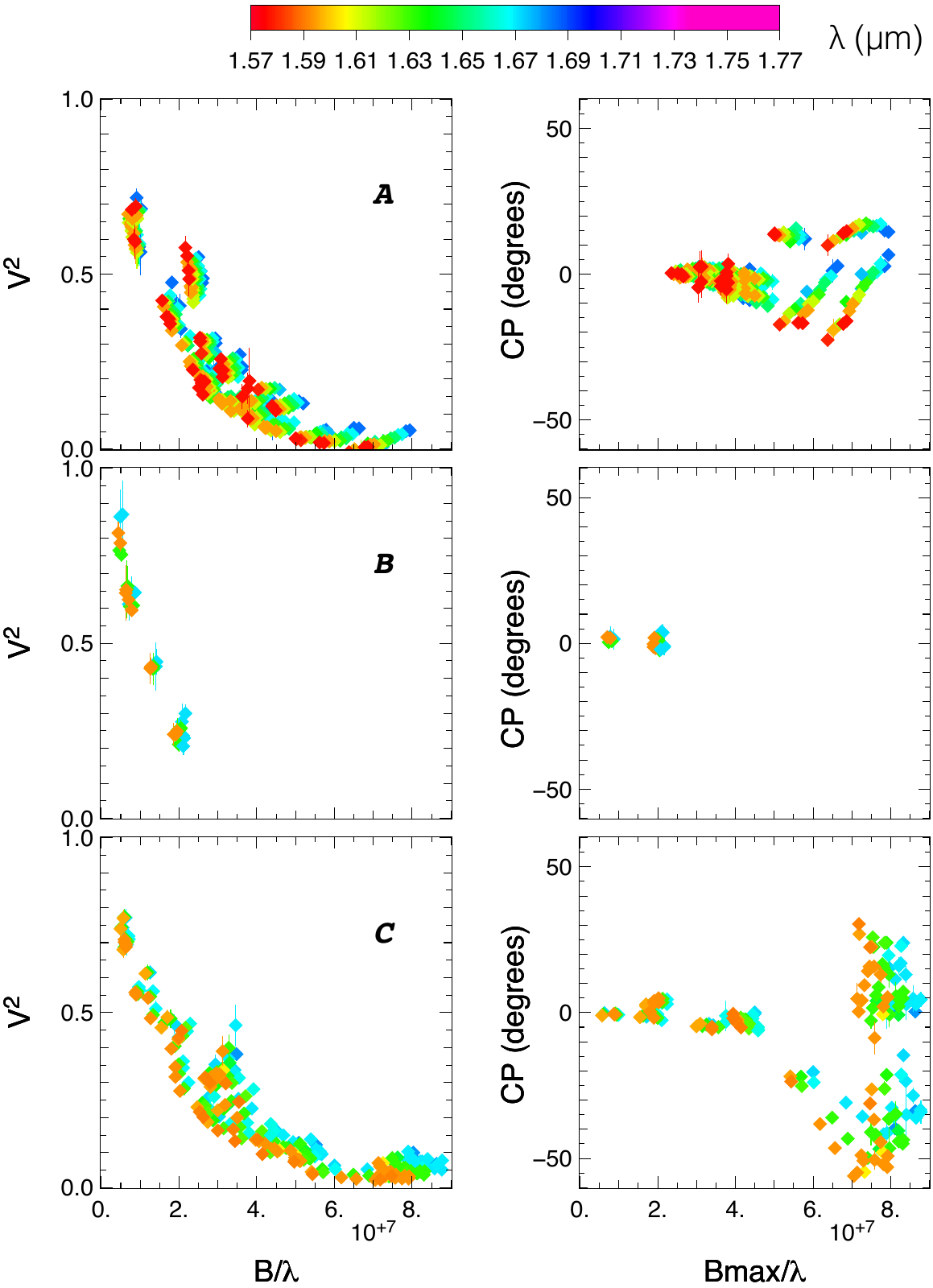} 
   \caption{Temporal representation on data taken on MWC158. The colours correspond to the wavelength. Dataset in V2 and CP. We notice strong differences in the \modif{CP} between the epochs that may indicate variability.}
   \label{fig:dataset}
\end{figure}

\modif{Epoch A was observed with the large spectral dispersion mode (seven channels across the $H$-band), while data of epochs B and C were observed with the small spectral dispersion mode (three channels across the $H$-band).}
The \uvs of the three epochs correspond to a maximum resolution of 2.6, 9.6, and 2.3\,mas, respectively, (with 72, 12, and 90 different baselines, respectively, and 504, 36, and 270 (u, v)-points, respectively) and are presented \modif{in} Fig.~\ref{fig:uv}. 
The associated dataset\modif{s} are presented \modif{i}n Fig.\,\ref{fig:dataset}.
A global V$^2$ decrease with spatial frequency indicates that the object is resolved.
The flat asymptote in V$^2$ at the highest spatial frequencies indicates that the object is sufficiently resolved to discriminate the stellar (point-like) flux from the CS environment. 
In parallel, we see that at a given baseline there is a decrease \modif{in} V$^2$ with increasing wavelengths.
This is because of the chromatic effect that arises when there is a temperature difference between the star and its CS environment \citep[][]{Kluska2014}.
For the smallest baselines, we see a drop \modif{in} V$^2$ that \modif{could} be because of the presence of an extended emission.

The \modif{CP} signal \modif{rises} significantly above 15\,degrees on epochs A and C, \modif{which} have sufficiently long baselines \modif{to resolve any departure from point symmetry in the \modiff{circum}stellar environment}.
\modif{The CP signal seems to be different from epoch to epoch.}

\section{Image reconstruction}
\label{sec:imgrec}

\begin{figure}[!t] 
   \centering
   \includegraphics[width=6cm]{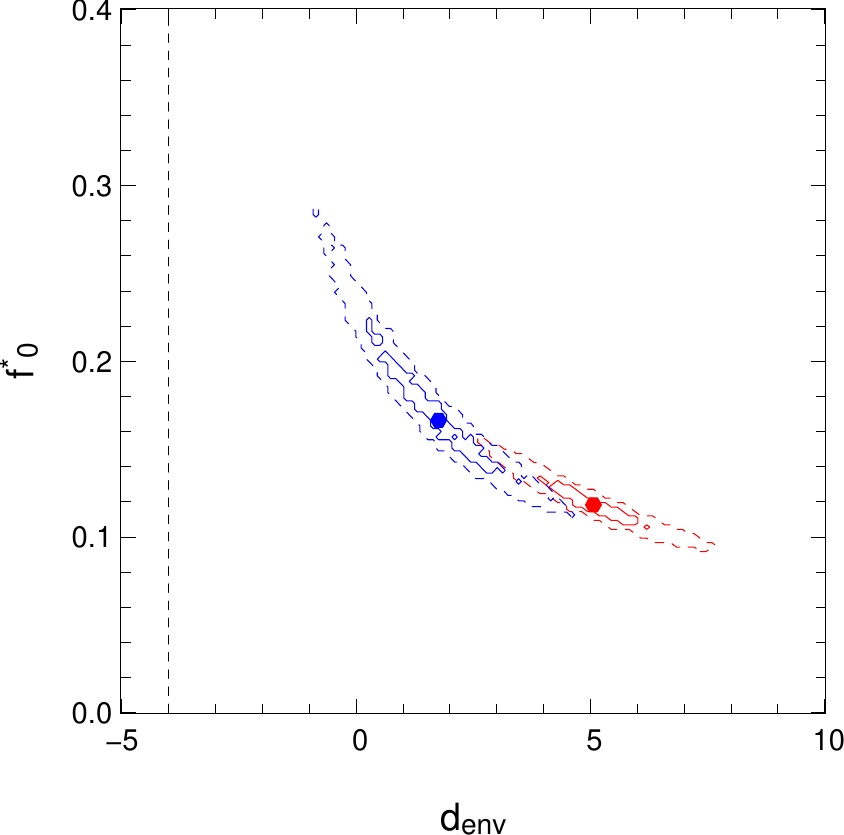} 
   \caption{Chromatic parameters determination for epoch A (red) and C (blue) at 1 $\sigma$ (solid lines) and 3 $\sigma$ (dashed lines). The red (epoch A) and blue (epoch C) points are the best values for the chromatic parameters. These values were used for the image reconstruction process. The vertical dashed line indicates the spectral index of the central star, \modif{which equals -4 because the star is assumed to be in the Rayleigh-Jeans regime}.}
   \label{fig:chrom}
\end{figure}

 We first use an image reconstruction technique to retrieve a model\modif{-}independent object morphology.
Because image reconstruction needs a sufficiently well covered \uv, we perform image reconstruction on the PIONIER datasets on epochs A and C only. 
The image reconstruction process requires a careful choice of the regularisation and its weight, which is described in Appendix~\ref{app:rgl}.

The images were made using the \texttt{SPARCO} approach \citep{Kluska2014} together with the \texttt{MiRA} algorithm \citep{Thiebaut2008}. 
In the SPARCO approach, the object is modelled as an unresolved central star plus an extended environment.
Both components have different spectral behaviours and therefore different spectral indexes.
The algorithm \modif{provides} an image of the extended environment for a given stellar-to-total flux ratio \modif{at 1.65$\mu$m}, \fso, and the spectral index of the environment, $\denv$ \citep[see Eq.~4 in][]{Kluska2014}.
It is possible to derive these chromatic parameters from the SED of the object but the variability of the object and the lack of simultaneous photometry prevent us from \modif{doing so}.
We therefore use the interferometric data itself to estimate these chromatic parameters. 
Several maps of the environment \modif{a}re reconstructed for a large number of  pairs of parameters, \fso and \denv, sampled on a 100$\times$100 grid (\fso from 0.05 to 0.25 for epoch A and 0.1 to 0.35 for epoch C, and \denv from 0 to 10 for epoch A and from -2 to 7 for epoch C).
We compute the \textit{a posteriori} image likelihood for each \modif{point in} this grid  (see App.\,\ref{app:rgl}).
We use different grids because we see that the ranges of chromatic parameters we need to sample are different.
The grids must be large enough to have negligible likelihood values at their edges.
We use bilinear interpolations to compare the grids.
We thus choose the optimal parameters with the highest \textit{a posteriori} image likelihood.
The result is plotted in Fig.~\ref{fig:chrom}.

 We use marginalised probability distributions to obtain the error bars on each parameter (see Fig.\,\ref{fig:marg} in the Appendix)\modif{.}
We notice that the result for the two epochs is different. 
Indeed, epoch A points to a value of the stellar-to-total ratio of \fso = 11.\modiff{9}\% $\pm 1.0$\% and a spectral index of the environment of \denv = 5.0 $\pm 0.8$ (T=872$^{+72}_{-65}$\,K), while epoch C gives \fso = 17.5\% $\pm 2.8$\% and a spectral index of the environment of \denv = 1.6 $\pm 0.9$ (T=1321$^{+209}_{-158}$\,K). 
These results confirm the variable nature of the object and imply a \modif{change in the} environment temperature of more than 200\,K.
The probability that the two epochs have the same chromatic parameters is less than 2\%.
Also, the stellar-to-total flux ratio is changing from 11.9\% to \modif{17.5}\% and agrees with the variability seen in photometric measurements (Lazareff, priv. comm.).
We want to investigate the origin of this variability by performing image reconstruction on both epochs.
We used the values with the highest probabilities \fso $=$ 11.8\% and \modiff{16.7}\% and $\denv = 5.1$ and $1.8 $ for epoch A and C, respectively.
To see the differences between the images at other flux ratios (\fso) and spectral indexes (\denv), we show image reconstruction with different pairs of chromatic parameters in the Appendix, Fig.\,\ref{fig:mos}.
No significant differences are seen in the image over the range of allowed chromatic parameters.
More importantly, the departures from point symmetry in the image are not affected.

\begin{figure*}[!t] 
   \centering
      \includegraphics[width=12cm]{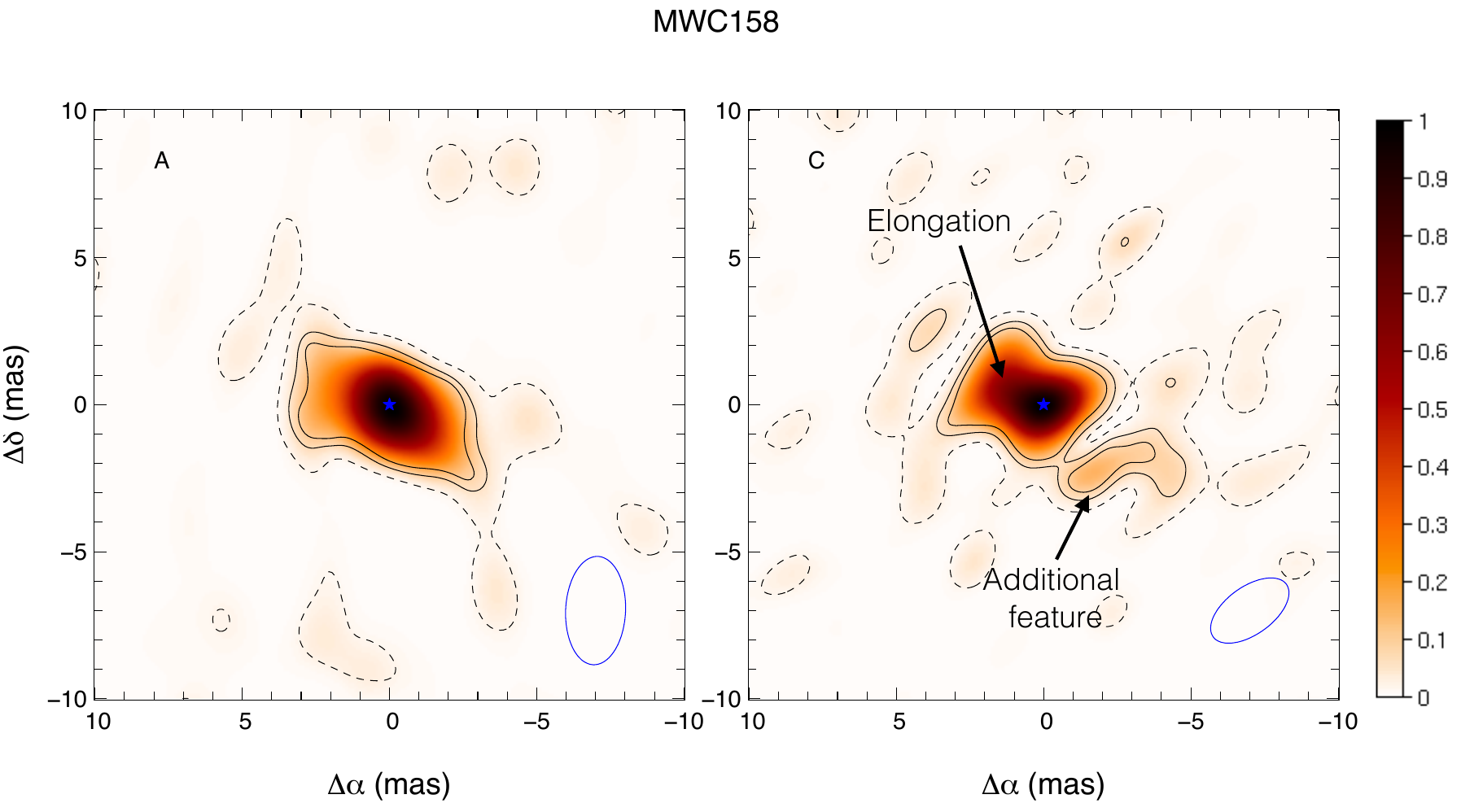} 
   \caption{Image reconstructions in $H$-band on PIONIER data, epochs A (left) and C (right). The colour bar on the right was redefined to have the maximum flux equal to one for the sake of readability. The beam for each epoch is plotted in blue in the bottom right corner of the images. The contours are at 1 $\sigma$ (dashed line), 3, and 5 $\sigma$ (solid lines) significance levels.}
   \label{fig:imgmwc}
\end{figure*}

\subsection{Image analysis}
\label{sec:boot}

We apply a bootstrap method \citep[e.g.][]{Boot} on the baselines and \modif{telescope triplets used to compute the CP} to determine the reliable features in the images.
\modif{This} consists \modif{of} a random draw of the data points within the whole dataset.
The generated dataset has the same number of points \modif{as} the original dataset.
Some data points can be drawn several times whereas some of them are not selected at all.
This allows the effects of an imperfect \uv to be revealed.
This process was carried out 500 times and led to a per-pixel error estimation.

The mean images are presented in Fig.\,\ref{fig:imgmwc} with contours at 1 (dashed line), 3 and 5 $\sigma$ (solid lines). 
The $\chi^2_\mathrm{red}$ of the images equal 2.0 for both epochs (see Fig.\,\ref{fig:qualimg} in the Appendix).

The two images reveal an elongated morphology, detected at more than 5 $\sigma$ with  a major axis of $\approx$ 6\,mas and a minor axis of $\approx$ 3 mas.
The PA of the major ax\modif{e}s are similar for the two epochs and are in good agreement with previous interferometric measurements \citep{BorgesII}.
The differences  mainly result from the relative positions between the star and the maximum brightness of the environment.
While in epoch A the image seems to be symmetric with respect to the star, this is not the case in epoch C.
At this epoch the environment is elongated in the north-east direction indicating a strong asymmetry detected at more than 20 $\sigma$.
Moreover, \modif{the} epoch C image has an additional feature appearing at the south-west direction, which is detected at \modif{greater} than 5 $\sigma$.

\subsection{Confirmation of the variability in Fourier space}
\label{app:compare}

\begin{figure*}[!t] 
   \centering
      \includegraphics[width=5cm]{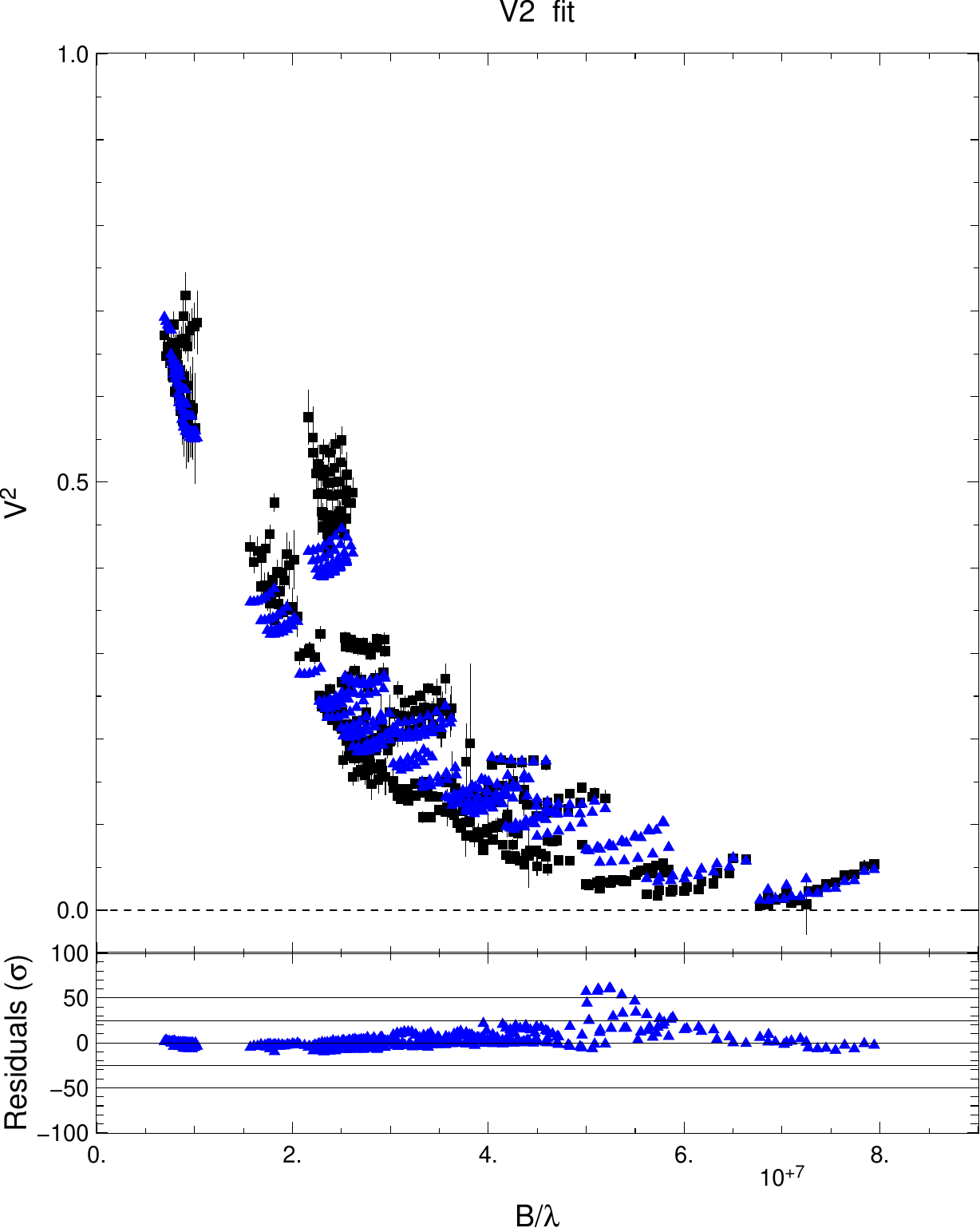} 
  \includegraphics[width=5cm]{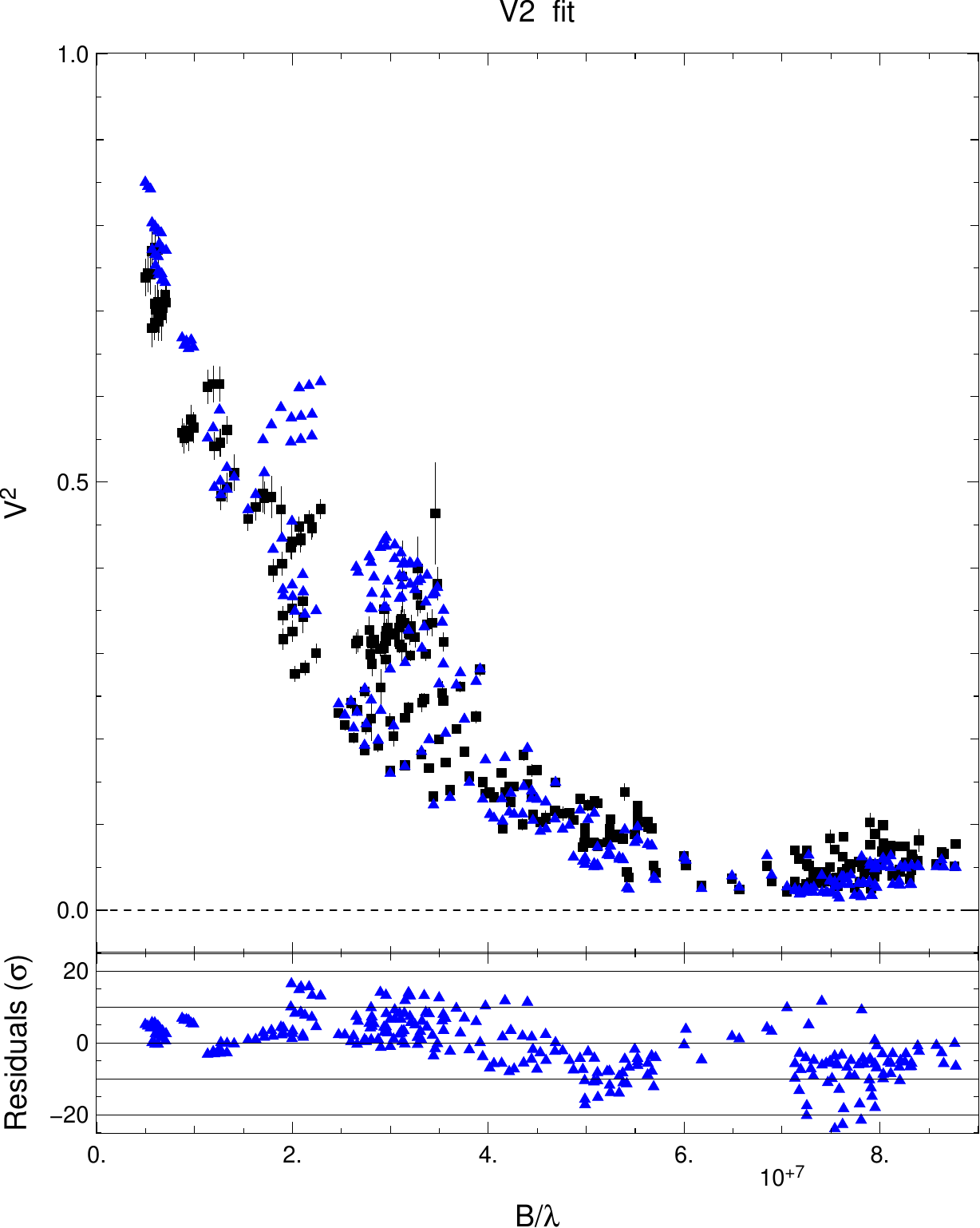} \\
   \includegraphics[width=5cm]{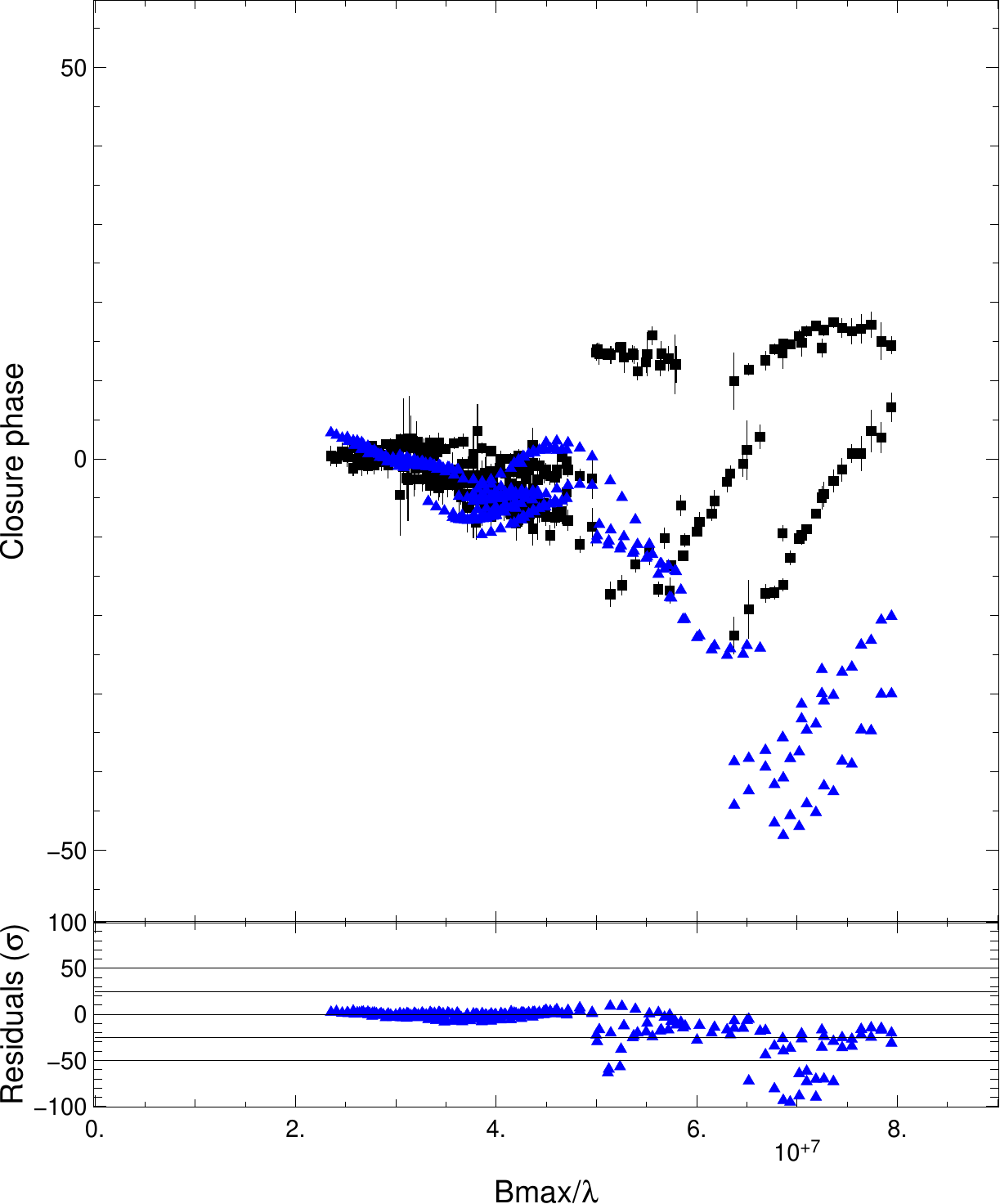} 
  \includegraphics[width=5cm]{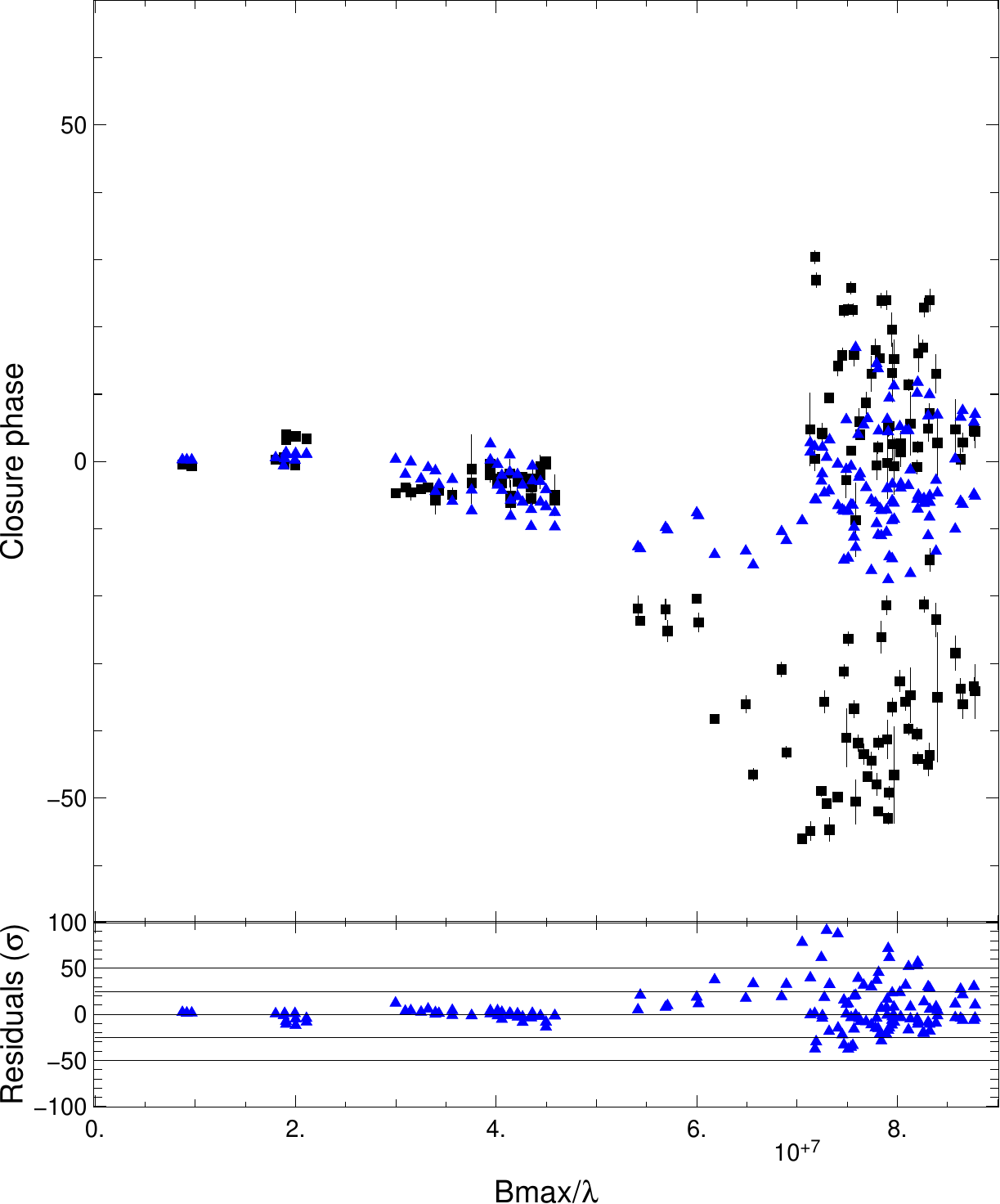}
   \caption{Image reconstruction fit on datasets of epoch C superimposed on dataset from epoch A (left) and inversely (right). The black squares represent the data and the blue triangles represent the images. The residuals normalised to the standard deviation are plotted in the bottom of each graph. }
   \label{fig:diff}
\end{figure*}
 
These two images appear very different  from each other. 
The differences arise from the environment brightness distribution and from the stellar-to-total flux ratio.
It is difficult to conclude the variability of an object if we do not measure precisely the same points in the \uv.
From the images we are able to extrapolate the values of the visibility through the entire Fourier plane.
We can therefore compare the visibilities from the image at one epoch to the dataset on the other epoch.
The superimposition of the data from the image of epoch C (A respectively) to the dataset of epoch A (C resp.) are in Fig.~\ref{fig:diff}.

We can see that the differences in the squared visibilities are \modif{generally} between 5 and 10 $\sigma$ (with peaks up to 60 $\sigma$).
The \modif{CP} differences are smaller at baselines \modif{shorter} than 80\,m that correspond to an angular resolution of 4\,mas.
At longer baselines the differences \modif{increase} significantly to 100 $\sigma$.
This is because of the elongations seen in the two \modif{epochs} that appear at sizes smaller than 4\,mas.
The variability of the environment is therefore highly significative.

\subsection{Asymmetry analysis}

\begin{figure*}[!t] 
   \centering
\includegraphics[width=12cm]{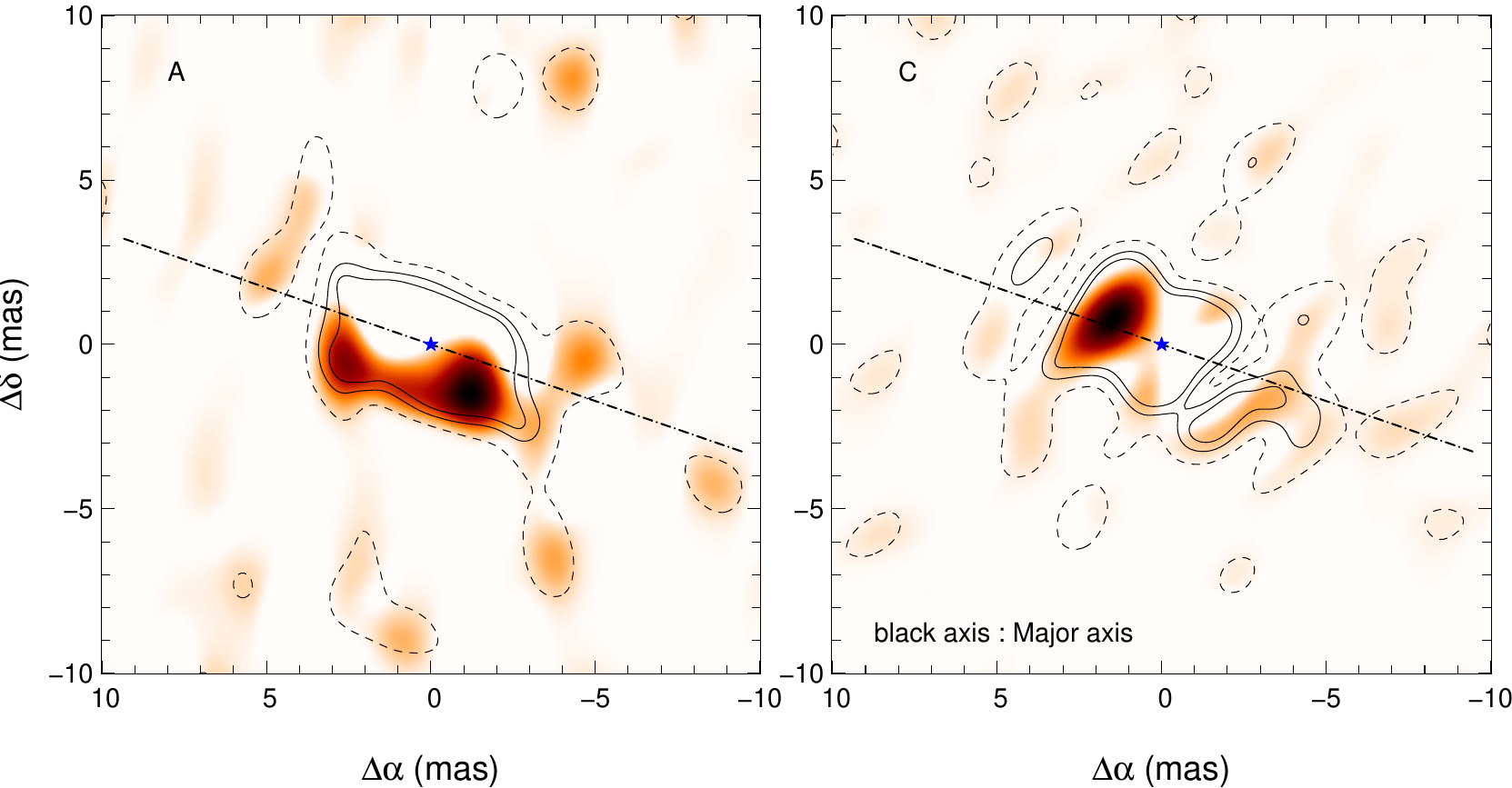} 
   \caption{Positive part of the asymmetric maps ($Asym$) of epochs A (left) and C (right). The orientation of the major axis \citep[P.A. of 71\,degrees as estimated in][]{BorgesII} is plotted in black in the image.}
   \label{fig:asym}
\end{figure*}

To emphasise the deviations from centro-symmetry in a model-independent way, we construct an asymmetric map ($Asym$).
It is obtained taking the positive part of
\begin{eqnarray}
Asym = \frac{1}{2} (Img - Img_{180})
,\end{eqnarray}
where $Img$ is the reconstructed image and $Img_{180}$ the same image rotated by 180 degrees.
\modif{The asymmetry map is the flux in the image that produces the non-zero \modif{CP}. 
It is the amount of flux that is not point-symmetric.}

\modif{In Fig.\,\ref{fig:asym} we plot the positive part $Asym$ for both epochs.
}
For epoch A, the positive parts of the asymmetry are on the southern side of the major axis \citep[P.A. of 71\,degrees as estimated in][]{BorgesII}. 
This is what is expected for an inclined inner disk rim, for instance, where we see the further side of the rim because the closer part is obscured by the disk itself. 
This is not the case in epoch C anymore where the position of the asymmetric flux is close to the major axis.
In both images the positions of the asymmetric flux seem to be at the same distance from the star.
That could be an indication of rotation of the asymmetric feature around the central star.
We use parametric models  to quantify this \modif{spatial} variability.

\section{Modelling the disk with asymmetric emission}
\label{sec:binfit}
The analysis of the reconstructed images at different epochs leads us to the hypothesis that there could be \modif{a moving} feature in the disk. 
We refer to this feature as \modif{a bright spot}.
We explore this hypothesis with a parametric model.
Because parametric models do not require a \uv that is as rich as for an image, we can divide the epochs into sub-epochs to better sample the trajectory of the \modif{asymmetry}.
We therefore divided the epochs A and C into the following sub-epochs (see Appendix\,\ref{app:log2}): A1 (1\,night), A2 (1\,night), A3 (2\,nights), C1 (2\,nights), C2 (3\,nights), C3 (3\,nights), and C4 (4\,nights).

\subsection{Model of a star, a disk, and \modif{an asymmetry}}
Our data supports the presence of a flattened morphology compatible with a disk in agreement with \citet{BorgesII}.
We built a \modif{simple} model of a star, disk, \modif{bright spot, and}  extended component.
The extended component can be interpreted as stellar light scattered by the disk \citep[see][]{Pinte2008,Anthonioz2015}.
From the image reconstructions we do not see any evidence for a disk inner rim.
We therefore modelled each component as a point source,  inclined Gaussian disk, background flux, and second point source that can be shifted w.r.t. the primary star. 
This model is detailed in Appendix\,\ref{app:mod2}.
\modif{All the flux ratios are derived for $\lambda_0 = 1.65\mu$m.}

We used a Levenberg-Marquardt method to find the best parameters. 
We start from 2000 different initial solutions generated by a Monte Carlo method to avoid local minima.
We estimate the errors on each parameter with the same bootstrap method already used in Sect.\,\ref{sec:boot}. 
We repeat it two thousand times in order to generate two thousand estimations of each parameter.
We use the standard deviation on each parameter to estimate its error (see Figure\,\ref{fig:paramerr}).


\subsection{Results}
\label{sec:paramfit}

\begin{table*}[!t]
 \caption{Best fits for each sub-epoch (A1, A2, A3, B, C1, C2, C3, C4)  with a star, Gaussian \modiff{disk} \modif{with an asymmetry} and extended component model. }
   \centering
   \begin{tabular}{@{} c|r|rr|rr|rr|rr|rr|rr|rr@{}} 
          & &\fso &  & $f^\mathrm{g}_0$ & & $f^\mathrm{bs}_0$ & &$T$ &(K) & $FWHM_\mathrm{g}$ & (mas) & $i_\mathrm{g}$ &(deg) &$PA_\mathrm{g}$ &(deg)\\
      \hline
      Epoch  & $\chi^2$ &Value & Err. & Value & Err. & Value & Err. &Value & Err. &Value & Err. &Value & Err. &Value & Err. \\
      \hline
A1&1.9 &26.4 & 0.6 &54.0 & 1.8 &4.8 & 0.4 &1445.6 & 97.9 &4.6 & 0.1 &60.2 & 1.7 &56.5 & 2.0  \\ 
A2& 2.8&33.6 & 2.4 &46.8 & 4.8 &8.0 & 2.4 &1987.2 & 246.5 &5.0 & 0.3 &55.6 & 3.5 &51.3 & 3.5 \\ 
A3& 6.1&18.5 & 0.3 &75.0 & 5.9 &2.9 & 0.1 &1298.2 & 127.5 &4.0 & 0.2 &54.3 & 4.8 &63.5 & 3.8 \\ 
B& 0.4 &42.4 & 4.9 &46.8 & 5.8 &1.0 & 0.6 &1508.0 & 429.3 &7.1 & 1.8 &64.3 & 21.6 &51.9 & 15.8  \\ 
C1&  0.7&35.1 & 3.2 &34.1 & 1.8 &20.5 & 2.7 &2417.1 & 483.4 &7.7 & 0.5 &64.5 & 13.4 &51.5 & 5.5 \\ 
C2& 8.4 &23.1 & 2.4 &25.6 & 6.8 &9.0 & 0.5 &1360.2 & 154.6 &2.4 & 0.3 &64.2 & 13.8 &78.8 & 14.6 \\ 
C3& 2.3&31.6 & 0.6 &41.6 & 6.5 &9.8 & 0.4 &1788.9 & 163.5 &4.3 & 1.1 &45.2 & 9.4 &88.0 & 37.3  \\ 
C4& 1.0&33.0 & 1.1 &47.5 & 2.0 &1.6 & 0.1 &1486.8 & 134.4 &4.4 & 0.1 &48.4 & 1.5 &65.4 & 1.7  \\ 
   \end{tabular}
   \tablefoot{Col. 1) \fso: stellar-to-total flux ratio; Col. 2) $f^\mathrm{g}_0$: inclined Gaussian \modiff{disk} flux ratio; Col 3) $f^\mathrm{bs}_0$: \modif{bright spot} flux ratio, Col. 4) T:  temperature of the Gaussian disk and \modif{bright spot}; Col. 5) $FWHM_\mathrm{g}$: full width at half maximum of the Gaussian disk, Col. 6) $i_\mathrm{g}$: inclination of the Gaussian \modiff{disk}; Col. 7) $PA_\mathrm{g}$:  position angle of the Gaussian \modiff{disk}. \modif{The flux ratios are derived for $\lambda_0 = 1.65\mu$m.} The extended-to-total flux ratio ($f^\mathrm{ext}_0$) can be deduced from the other flux ratios as $f^\mathrm{ext}_0 = 1 - \fso - f^\mathrm{a}_0 - f^\mathrm{g}_0$.}
      \label{tab:fit}
\end{table*}

\begin{table}[t]
 \caption{Continuation of Table\,\ref{tab:fit} with coordinates of the \modif{bright spot} $x_\mathrm{a}$ and $y_\mathrm{a}$}
   \centering
   \begin{tabular}{@{} c|r|rr|rr@{}} 
           & & $x_\mathrm{bs}$ & (mas) &$y_\mathrm{bs}$ &(mas) \\
      \hline
      Epoch  &$\chi^2$&Value & Err.&Value & Err. \\
      \hline
A1&1.9 & 1.9 & 0.1 &-1.6 & 0.3 \\ 
A2& 2.8& 2.1 & 0.1 &0.3 & 0.3 \\ 
A3& 6.1& -4.7 & 0.1 &-0.1 & 0.1 \\ 
B& 0.4 &-18.5 & 3.1 &19.4 & 2.7 \\ 
C1&  0.7 &3.0 & 0.4 &-1.2 & 0.3 \\ 
C2& 8.4 &1.6 & 0.1 &0.7 & 0.1 \\ 
C3& 2.3 &1.4 & 0.1 &0.7 & 0.1 \\ 
C4& 1.0 &-3.4 & 0.1 &-3.3 & 0.1 \\ 
   \end{tabular}
   \label{tab:fit2}
\end{table}

The best-fit parameters are presented in Tables\,\ref{tab:fit} and \ref{tab:fit2}. 
We find different $\chi^2_\mathrm{red}$ values for the different epochs.
Interestingly, the highest $\chi^2_\mathrm{red}$ correspond to the observations with the highest spatial resolution (epochs A3 and C2).
This indicates \modif{that our model is too simple and} that we are still not catching all the complexity of the object \modif{on} the smallest spatial scales.

We discuss the results of the fits for each parameter:
\begin{itemize}
\item \textbf{Flux of the \modif{asymmetry}: } At epoch B, the flux is not significant enough (less than 2 $\sigma$ significance) because it is not resolved. For all the other epochs, the \modif{asymmetry} has a flux that \modif{varies} from 1.6\% (epoch C4) to 20.5\% (epoch C1).
\item \textbf{Flux of the star: } The stellar flux \modif{ratio} is oscillating around 30\% with the highest value at epoch B (42.4\%) and the lowest values at epochs A3 and C2. We notice that the epochs with the highest values have short baselines and the epochs with the lowest values have the longest baselines. The estimation of the stellar flux is depend\modif{e}nt on the angular resolution achieved with respect to the size of the environment. If the angular resolution is sufficient to \modif{resolve the environment without resolving the star}, then the stellar-to-total flux ratio is unbiased. \modif{In contrast, the flux of the central star can be considered the unresolved flux and include unresolved disk flux. Having simultaneous photometry would have helped to isolate the stellar contribution in the unresolved flux and claim a possible continuum variability.}
\item \textbf{Flux of the Gaussian: } The flux of the Gaussian that is representing the disk also varies. It seems to be correlated to the stellar flux variation for the epochs from B to C4. Interestingly the fluxes for epochs A1, A2 and A3 seem to be anti-correlated.
\item \textbf{Temperatures: } The temperatures found for \modif{the disk} are around 1400\,K.
We also notice that the temperatures are higher for epochs A2 and C1, which correspond to the observations with the less extended configurations, \modif{for which the environment is not entirely resolved. \modiff{In these cases, the unresolved flux is also included in the stellar flux ratio (\fso), thus the flux of the star cannot be considered anymore as in the Rayleigh-Jeans regime, as previously assumed, causing an overestimation of the disk temperature}}. These temperatures are quite different from those found by image reconstruction, especially for epoch A. 
\item \textbf{Size of the Gaussian: }The size of the Gaussian is stable (around 4.5\,mas) but for epochs \modif{B, C1 and C2}. Epochs \modif{B and C1} ha\modif{ve} the smallest angular resolution and it can affect the size determination. The best model for C2 epoch has the highest $\chi^2$ and is not reproducing the data very well. \modif{We do not therefore consider the size variation as significative.}
\item \textbf{Disk orientations: } The inclination of the disk is consistent through the epochs with 60\,degrees (a bit lower for epochs C3 and C4). Its position angle is consistent with a value between 55 and 60\,degrees.
\item \textbf{\modif{Bright spot} coordinates: } The positions of the \modif{bright spot} are different from epoch to epoch \modif{but they remain close to the disk}.
The bright spot seems to follow a trajectory \modif{around the central star} that has the same orientation as the disk. \modif{It could therefore be a disk feature.}
\end{itemize}

The variation of the stellar flux ratio, the disk temperature, and size seems to be depend\modif{e}nt on the baseline length.
This is due to a degeneracy.
In our dataset the \uv is not sufficient to resolve all \modif{of} the \modiff{circum}stellar environment, the unresolved flux is therefore composed of the stellar flux and an unresolved \modiff{circum}stellar environment flux.
In the fitting process the stellar-to-total flux ratio (\fso), which is modelled by the unresolved stellar component, is therefore over estimated.
Moreover the assumption that the model of the unresolved star is in the Rayleigh-Jeans regime is \modif{no longer valid}.
As the spectral index of the environment is defined relative to the index of the star, it is \modif{not properly estimated}.
As a result the temperature derived from it will be overestimated (for ex. 1987\,K and 2417\,K for epochs A2 and C1 respectively).

\section{Discussion}
\label{sec:dis}

A physical interpretation \modif{of} the detected morphological variability is difficult from \modif{our} interferometric dataset \modif{alone} because of the insufficient temporal sampling of \modif{our} observations, however, we present some possible interpretations.

\subsection{A binary?}

\begin{figure}[!t] 
   \centering
   \includegraphics[width=6cm]{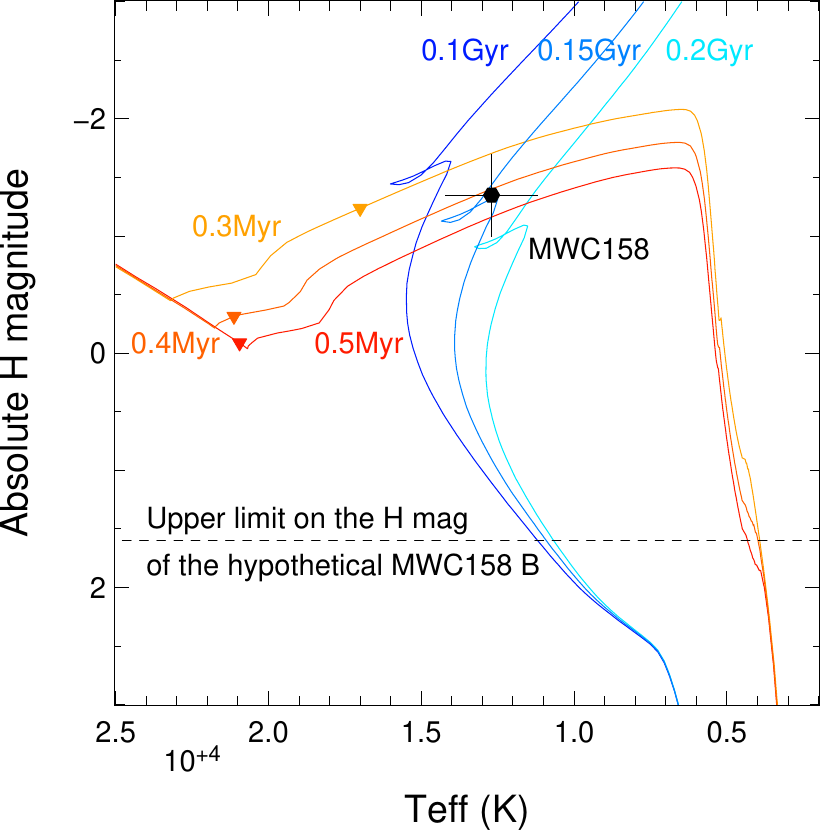} 
   \caption{\modif{Location of MWC158 primary star \modiff{(black hexagon, see text for details)} in the isochrones of \citet{Bressan2012} for two scenarios: the scenario in which MWC158 is young, from yellow (0.3\,Myr) to red (0.5\,Myr), and old, from blue (0.1\,Gyr) to cyan (0.2\,Gyr). The triangles indicate the position of 6 M$_\odot$ as determined by \citet{BorgesI} for the young track. For the old track, these positions are out of range of the plot. The dashed line indicates the upper limit on the absolute H magnitude of a hypothetical secondary of MWC158.}}
\label{fig:HRdiag}
\end{figure}

A binary can easily reproduce such a strong \modif{CP} signal with \modif{rapid} variability,
however all \modif{of} our binary fits alone (i.e. without a disk) are not able to reproduce the data ($\chi^2$ of 15 to 250 for the different sub-epochs). 
Previous spectroscopic observational campaigns do not show any binary signal \citep{Corporon_1999, BorgesI, BorgesIII} but they do not discard this possibility.
\modif{From the parametric fit in Sect.\,\ref{sec:paramfit}, we saw that a fit with a star, disk, and bright spot is able to reproduce correctly most of the dataset, especially the closure phases.}

\modif{With the flux ratios from the parametric fits, we can locate the primary star and the asymmetry (if related to a secondary) on a Hertzprung-Russel diagram.
For the primary we used the effective temperature range derived by \citet[11200\,K \textless Teff \textless 14200\,K]{BorgesI} and the absolute $H$-magnitude range derived from the \fso \modif{by our modelling.}
From \modiff{isochrones of} \citet{Bressan2012}, we can see \modiff{the presence of two scenarios for the primary star of MWC158} (see Fig.\,\ref{fig:HRdiag}).
A ``young" scenario with ages from 0.3 to 0.5\,Myr and an ``old" scenario with ages from 0.1 to 0.2\,Gyr.
The masses associated with the ``young" scenario range from 4.6 to 5.7\,M$_\odot$ whereas the ``old" scenario predicts masses from 3.7 to 4.8\,M$_\odot$.
These masses \modiff{are in agreement with the literature} \citep{BorgesI}.
}
Assuming that the flux of the \modif{asymmetry} is composed of the flux of the secondary and from surrounding parts of the disk, we can estimate an upper limit for the secondary mass.
If we take the minimum flux value for the \modif{asymmetry}, which is 5\% of the flux of the primary for epoch C4 (that corresponds to an absolute magnitude in $H$ of \modif{1.6}), we can estimate the upper mass limit of the secondary using model isochrones.
\modif{Assuming that the secondary would have the same age as the primary and that no mass exchange happened, the secondary would have a maximum mass of 0.6\,M$_\odot$ in the ``young" scenario and of 2.3\,M$_\odot$ in the ``old" scenario.}
\modif{The total mass of the binary system would be about 7\,M$_\odot$.}
\modif{With such a mass and taking an orbital radius of  4.7\,mas (the position of the spot of the C4 epoch), we expect a Keplerian period of almost one year (344\,days).
Looking at the sequence of C1, C2, C3, and C4 epochs that lasts two months, however, we can see that the asymmetry changes its position \modiff{on a shorter} timescale than our prediction (we find the same conclusion for epochs A1, A2, and A3).}

\modif{We notice that the coordinates of the asymmetry are distributed around the star and may trace an orbital trajectory around it.}
\modif{We tried to fit a circular orbit (see Appendix\,\ref{app:bin}) but it was unsuccessful. }
\modif{The trajectory of the spot could be more complex than a circular Keplerian orbit but we do not have good enough temporal sampling to characterise it more accurately.}
\modif{Another possibility is that our parametric model is too simple to recover all the complexity of the environment morphology.}

\subsection{An asymmetric environment ?}

\modif{From the parametric fit the positions of the bright spot are located close to the disk and this can be interpreted as an asymmetry lying in the disk.}
\modiff{This is supported by a spectroscopic campaign over 20 years on this object, which provides evidence for an asymmetric gaseous disk; this was carried out by \citet{Jerabkova2015}.}
\modiff{The authors could} not find a clear periodical behaviour but they claimed several pseudo-periods ranging from years to months.

\modiff{Several scenarios could explain such asymmetric features.}
\citet{BorgesI} postulated \modiff{the formation of a "one-armed spiral" due to matter ejection from a stellar hot spot.} 
\modiff{Asymmetric features were also observed in the circumstellar environment of the AGB star L$_2$ Puppis \citep{Kervella2014,Lykou2015}.
Such a pattern is very likely generated by the interaction of a close companion (that we could not constrain around MWC158) with the strong stellar wind coming from the primary \citep{vanWinckel2009}.}

\modif{Another possible \modiff{explanation for the observed asymmetry} lies in shell ejection episodes.}
The photometric variability of this object has led several authors to postulate episodes of shell ejection for this star \citep{Merrill1931,Merrill1933,Doazan1965,Hutsemekers1985,Andrillat1991,Bopp1993,Pogodin1997,BorgesI,BorgesIII,Jerabkova2015}.
\modiff{The asymmetry could be created by these ejections that would carry a part of the gas and dust away.}
Such a phenomenon was already proposed by \citet{Bans2012} with a disk wind model.
The asymmetric shell ejection scenario could explain the fact that at epoch B the object was more extended that in epochs A and \modif{C}.
In order to reproduce our data the shell should be very asymmetric and its brightness cannot only evolve radially.
In order to test such a possibility, the morphological variability of the environment of this object should have been observed more extensively to achieve a better temporal sampling.


\section{Conclusions}
\label{sec:ccl}
The aim of these observations was to constrain the morphology of the environment of MWC158.
The orientations and sizes that we derived are in clear agreement with what was found in previous studies \citep{BorgesII}.
We reconstructed two images at different epochs.
For the first time we have detected a strongly asymmetric environment that is \modif{spatially} evolving with time.
We modelled the asymmetric environment with a simple model of a star, a disk and a bright spot.
We found that the \modif{bright spot} has different coordinates from epoch to epoch and seems to move around the central star.
The \modif{result} of \modif{fitting} a circular orbit is inconclusive.
It is either due to the lack of additional epochs or to the fact that the object is more complex than our modelling, as already suggested \citep{BorgesI,Jerabkova2015}.
Because it is located close to the star, it is likely that the bright spot is a disk asymmetry, however, the detected asymmetry cannot be linked directly with the presence of a secondary.
It could be caused by spiral arms lying in the disk \citep{BorgesI,Jerabkova2015} or other disk features.

We do not exclude the scenario of an asymmetric shell ejection that could explain this data as well, but it needs to be further explored.
The origin of the asymmetry could also be partly due to the circumstellar gas or an additional stellar companion that is not confirmed in our dataset.
More observations with better temporal sampling are needed to assess the displacement of the asymmetry.

The environment of MWC158 cannot be considered as axisymmetric and static.
Simultaneous interferometric observations of the continuum and circumstellar lines are necessary to constrain the dynamics in place in this object.
This could be possible in the context of the Gravity instrument by observing the Br$_\gamma$ line \modif{and the continuum} simultaneously.
Spatially resolving the kinematics will indicate whether the inner gaseous disk is disrupted in the same way as the dusty continuum.
Nevertheless, to understand the variability of the environment, simultaneous spectrophotometric monitoring is needed to follow up the stellar-to-total flux ratio to correctly interpret the interferometric datasets.
It would also be exciting to constrain the dust at larger scales, in order to detect the possible ejection shells or spirals caused by an inner companion, with direct imaging of the scattered light close to the star in the visible, for instance, with SPHERE/ZIMPOL.

In the context of B[e] stars, MWC158 is a key case because it is one of the brightest stars showing this phenomenon and it is relatively close.
If the B[e] phenomenon is linked with ejections without being a binary, it will bring new elements to explain some stars showing the B[e] phenomenon.

\begin{acknowledgements}
We thank the anonymous referee for remarks that greatly improved the manuscript.
We would like to thank Stefan Kraus for fruitful discussions.
This work is supported by the French ANR POLCA project (Processing of pOLychromatic interferometriC data for Astrophysics, ANR-10-BLAN-0511).
J.K. acknowledges support from an Marie Sklodowska-Curie CIG grant (Ref. 618910, PI: Stefan Kraus).
FS was partly supported by the Sinergia project  ``Euclid"  from the Swiss National Science Foundation.
PIONIER is funded by the Universit\'e Joseph Fourier (UJF), the Institut de Plan\'etologie et d'Astrophysique de Grenoble (IPAG), the Agence Nationale pour la Recherche (ANR-06-BLAN-0421, ANR-10-BLAN-0505, ANR-10-LABX56), and the Institut National des Science de l'Univers (INSU PNP and PNPS). The integrated optics beam combiner is the result of a collaboration between IPAG and CEA-LETI based on CNES R\&T funding. The authors want to warmly thank all
the people involved in the VLTI project. This work is based on observations made
with the ESO telescopes. It made use of the Smithsonian NASA Astrophysics
Data System (ADS) and of the Centre de Donn\'ees astronomiques de Strasbourg
(CDS). All calculations and graphics were performed with the open source software
\texttt{Yorick}. This research has made use of the Jean-Marie Mariotti Center ASPRO2 and SearchCal services  co-developed  by  CRAL,  IPAG,
and  FIZEAU. 

\end{acknowledgements}


\appendix

\section{Log of observations}
\label{app:log2}
The log of the observations is in Table~\ref{tab:log2}.

\begin{table*}[ht]
\caption{Log of observations with PIONIER/VLTI}
\begin{center}
\begin{tabular}{ccccccc}
Epoch &  & Date & mjd & Station index & Baseline (m) & PA ($\deg$)\\
\hline
 A&A1&2\modiff{7}/10/2010&55496.3&D0-E0-H0-I1&58.5-67.9-12.1-40.4-48.7-36.5&166.6-173.6-29.6-128.5-29.6-29.6\\ 
 &&2\modiff{7}/10/2010&55496.3&D0-E0-H0-I1&62.8-13.3-40.7-40-73.6-53.4&166.6-25.9-128-25.9-173.2-25.9\\ 
 &&2\modiff{7}/10/2010&55496.3&D0-E0-H0-I1&13.5-40.7-40.5-63.4-74.4-54&25.4-127.9-25.4-166.5-173.1-25.4\\ 
 &&2\modiff{7}/10/2010&55496.3&D0-E0-H0-I1&75.2-64-13.6-41-54.7-40.7&173-166.5-24.9-24.9-24.9-127.8\\ 
 &&2\modiff{7}/10/2010&55496.3&D0-E0-H0-I1&14.6-66.9-79.2-58.7-40.6-44&22.1-166.1-172.3-22.1-126.5-22.1\\ 
 &A2&01/12/2010&55531.2&E0-G0-H0-I1&65.4-40.7-42.4-54.8-14.1-28.2&166.4-127.3-23.6-157.4-23.6-23.6\\ 
 &&01/12/2010&55531.2&E0-G0-H0-I1&67.9-45.3-15.1-56.2-30.2-40.4&165.7-20.9-20.9-156.8-20.9-125.5\\ 
 &&01/12/2010&55531.3&E0-G0-H0-I1&47.1-31.4-68.7-39.8-56.4-15.7&19-19-164.8-123.2-155.8-19\\ 
 &&01/12/2010&55531.3&E0-G0-H0-I1&38.9-68-47.9-31.9-55.5-15.9&120.1-163.6-17.6-17.6-154.3-17.6\\ 
 &A3&21/12/2010&55551.3&A0-G1-I1-K0&45.9-46.3-90.4-72.5-92.4-114.6&43.6-66.3-55-143.7-172.9-15.7\\ 
 &&22/12/2010&55552.3&A0-G1-I1-K0&80.9-103.3-123.7-45.7-46.6-90.1&149.3-175.1-16-69.5-44.2-56.8\\ 
 &&22/12/2010&55552.3&A0-G1-I1-K0&45.8-122.1-101.3-79.2-46.5-90.3&68.7-15.8-174.7-148.2-44-56.3\\ 
 \hline
 B&&27/04/2012&56044.0&A1-B2-C1-D0&33.8-8.6-13.7-11.2-22.5-34.4&54.7-141.1-15.8-54.7-54.6-40.1\\ 
 &&27/04/2012&56044.0&A1-B2-C1-D0&32.9-11.1-12.3-22.3-7.7-33.4&41.1-54.5-16.5-54.4-134.3-54.4\\ 
 \hline
 C&C1&19/12/2012&56280.2&A1-B2-C1-D0&21.2-33.6-31.8-11.3-15.4-10.6&67.1-47.6-67.2-156.6-20-67.2\\ 
 &&19/12/2012&56280.3&A1-B2-C1-D0&35.7-10.5-22.4-33.5-11.1-15.7&40.9-151.3-58-58-58-16.3\\ 
 &&20/12/2012&56281.1&A1-B2-C1-D0&9.9-29.8-19.9-8.7-27.1-9.9&91.4-91.3-91.3-155.2-74.6-39.3\\ 
 &C2&26/01/2013&56318.3&A1-G1-J3-K0&47.2-127.1-96.3-132.2-87.9-49.5&140-45.6-25.5-66.5-54.7-87.7\\ 
 &&27/01/2013&56319.2&A1-G1-J3-K0&51.1-139.6-90.3-70-129.2-124.4&106.3-43.9-56.1-157.1-73.8-22.5\\ 
 &&27/01/2013&56319.2&A1-G1-J3-K0&61.7-136.6-115.8-130.7-90.2-49.9&152.2-43.6-22.8-70.2-54.7-99.1\\ 
 &&27/01/2013&56319.3&A1-G1-J3-K0&89.6-133.5-109-56.2-131.5-49.6&54.4-44.1-23.5-148.4-68.5-94.7\\ 
 &&28/01/2013&56320.3&A1-G1-J3-K0&132.2-49.5-87.7-46.6-95.3-126.6&66.4-87.2-54.8-139.2-25.6-45.8\\ 
 &C3&30/01/2013&56322.1&A1-G1-J3-K0&54.6-79.8-135.5-86.7-127.3-126.1&118.7-163.1-49-63.4-25.3-84.3\\ 
 &&31/01/2013&56323.1&A1-G1-J3-K0&79.9-55.1-125.9-133.8-85.9-125.8&163.6-120-26-50.3-65-86\\
 &&01/02/2013&56324.1&A1-G1-J3-K0&78.6-53.7-126.7-88-137.8-128.8&162-116.1-81.3-60.9-47.1-24.1\\ 
 &&01/02/2013&56324.2&A1-G1-J3-K0&139.8-127.8-89.6-128-74.6-52.2&44.8-22.9-57.8-76.8-159.6-110.9\\
 &C4&1\modiff{7}/02/2013&56340.1&D0-G1-H0-I1&73.5-61-36.6-46.5-62.8-69.9&166-15.8-110.2-44-127.1-72.7\\
 &&\modiff{18}/02/2013&56341.2&D0-G1-H0-I1&44.8-56.7-35.2-71.1-53.1-60.1&43.9-115.4-99.2-67.9-16-160.4\\ 
 &&\modiff{20}/02/2013&56343.2&D0-G1-H0-I1&48.1-52.7-54.4-34.9-43.4-71.4&16.7-156.6-108.8-93.8-44.7-66.4\\
\end{tabular}
\end{center}
\label{tab:log2}
\end{table*}%

\section{Model of a star, disk with a "bright spot", and an extended component}
\label{app:mod2}
The star has a visibility of 1.
The extended flux has a visibility of 0.
The inclined Gaussian is geometrically described by its \modiff{FWHM} ($w$), its inclination ($i$), position angle ($PA$) as follows:
\begin{eqnarray}
V^\mathrm{g} (\nu') &=& \exp\Big(-\frac{(\pi w_1 \nu')^2}{4 \log2}\Big)
,\end{eqnarray}
where $V^\mathrm{g} (\nu')$ is the visibility of the Gaussian and $\nu' = \sqrt{u'^2+\mathrm{v}'^2}$ are the spatial frequencies from the \uv points oriented with respect to the object such as 
\begin{eqnarray}
u' &=& u \cos PA + \mathrm{v} \sin PA\\
\mathrm{v}' &=& (-u \sin PA + \mathrm{v} \cos PA) \cos i
,\end{eqnarray}
with $\nu=\sqrt{u^2 + \mathrm{v}^2}$ the spatial frequencies from the \uv.

The bright spot is mainly described by its position $x_{bs}$ and $y_{bs}$ as follows: 
\begin{eqnarray}
V^\mathrm{a} (\nu'') & = & \exp(-2 i \pi \nu")
,\end{eqnarray}
where $V^\mathrm{a} (\nu'')$ is the visibility of the bright spot and $\nu" = u"+\mathrm{v}"$ are the spatial frequencies from the \uv points shifted with respect to the object such as \begin{eqnarray}
\mathrm{u}" &=& \mathrm{u} x_{bs}  \\
\mathrm{v}" &=& \mathrm{v} y_{bs}
.\end{eqnarray}

The total visibilities are
\begin{eqnarray}
V^\mathrm{tot}(\nu, \lambda) = \frac{f^*_{\lambda} + f^\mathrm{g}_{\lambda} V^\mathrm{g} (\nu) + f^\mathrm{bs}_{\lambda} V^\mathrm{bs} (\nu)}{f^*_{\lambda} + f^\mathrm{g}_{\lambda} +f^\mathrm{a}_{\lambda} + f^\mathrm{ext}_{\lambda}}
,\end{eqnarray}
with $f^*_{\lambda}$ the stellar to total flux ratio, $f^\mathrm{ext}_{\lambda}$ the extended-to-total flux ratio, $f^{g}_{\lambda}$ the Gaussian-to-total flux ratio, and $f^{bs}_{\lambda}$ the bright spot-to-total flux ratios.
To define these parameters, we have taken a reference wavelength $\lambda_0 = 1.65\mu$m (centre of the $H$-band). 
We assumed the star ($f^*_{\lambda}$) and the extended flux ($f^\mathrm{ext}_{\lambda}$) being in the Rayleigh-Jeans regime ($f^*_{\lambda} = \fso (\frac{\lambda}{\lambda_0})^{-4}$ and $f^\mathrm{ext}_{\lambda} = f^\mathrm{ext}_0 (\frac{\lambda}{\lambda_0})^{-4}$) and that the Gaussian and the bright spot ($f^\mathrm{g}_{\lambda}$, $f^\mathrm{bs}_{\lambda}$) have their own temperatures so that their fluxes are scaled accordingly and $f_{\lambda} = f_0 \frac{\mathrm{B}(\lambda,T)}{\mathrm{B}(\lambda_0,T)}$. 
We assumed the temperature of the Gaussian and the bright spot to be equal.
At $\lambda_0$, the sum of the stellar flux ratio (\fso),  extended flux ratio ($f^\mathrm{ext}_0$)\modiff{,} inclined Gaussian flux ratio ($f^\mathrm{g}_0$), and bright spot ($f^\mathrm{bs}_0$) equals unity.

~

\section{Parameter error determination}
\label{app:err}

In order to estimate the errors on the parameters from the parametric fit, we have generated 2000 new datasets from the original one adding a noise term. We assumed a Gaussian noise with a standard deviation equal to that from the data.
We then have computed the mean value and the error on each parameter from the best fit to the generated dataset.
The standard deviation for each parameter gave us the error at 1 $\sigma$ that we have used.
~

\begin{figure*}[!t] 
   \centering
   \includegraphics[width=8cm]{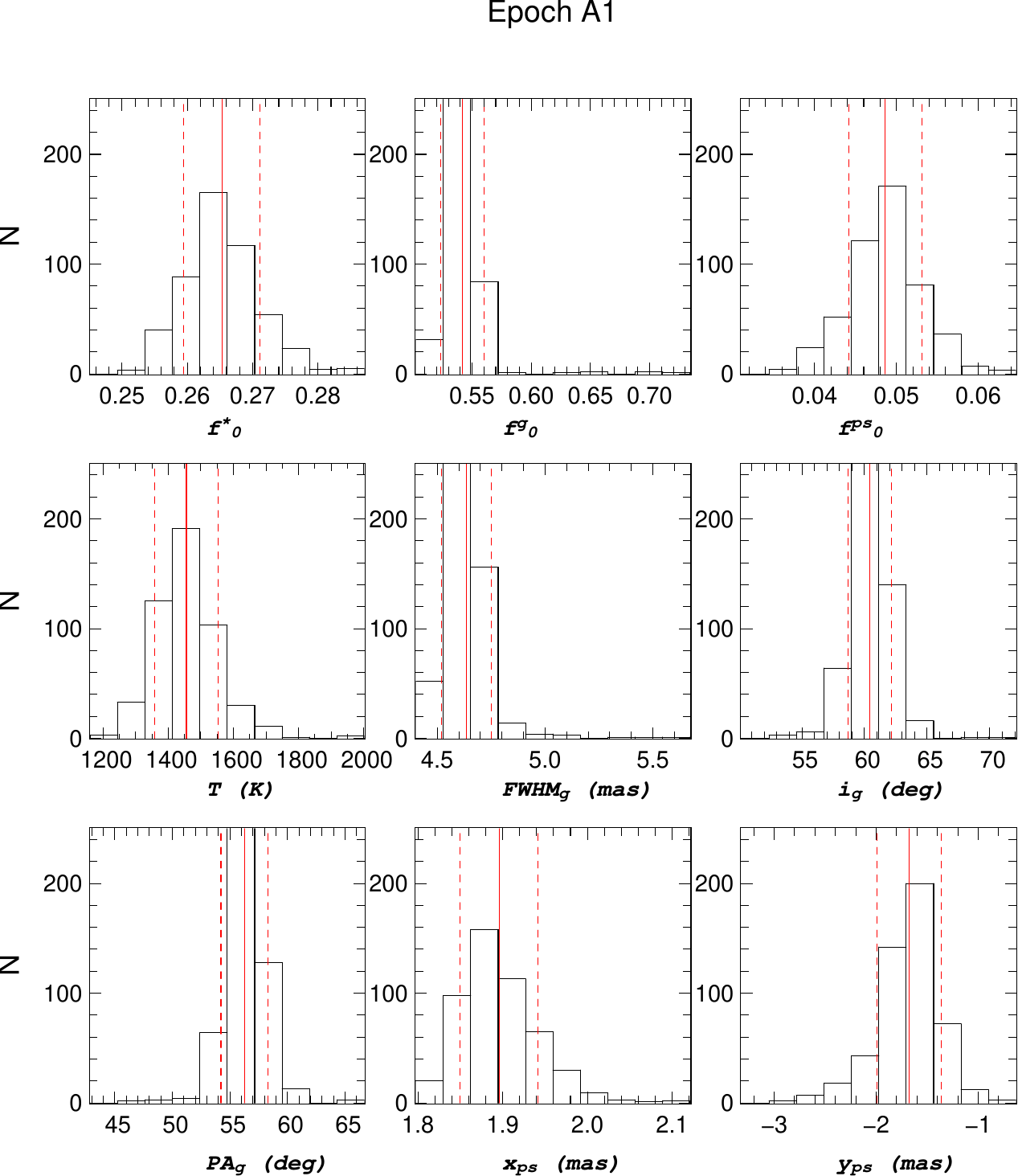} 
 \includegraphics[width=8cm]{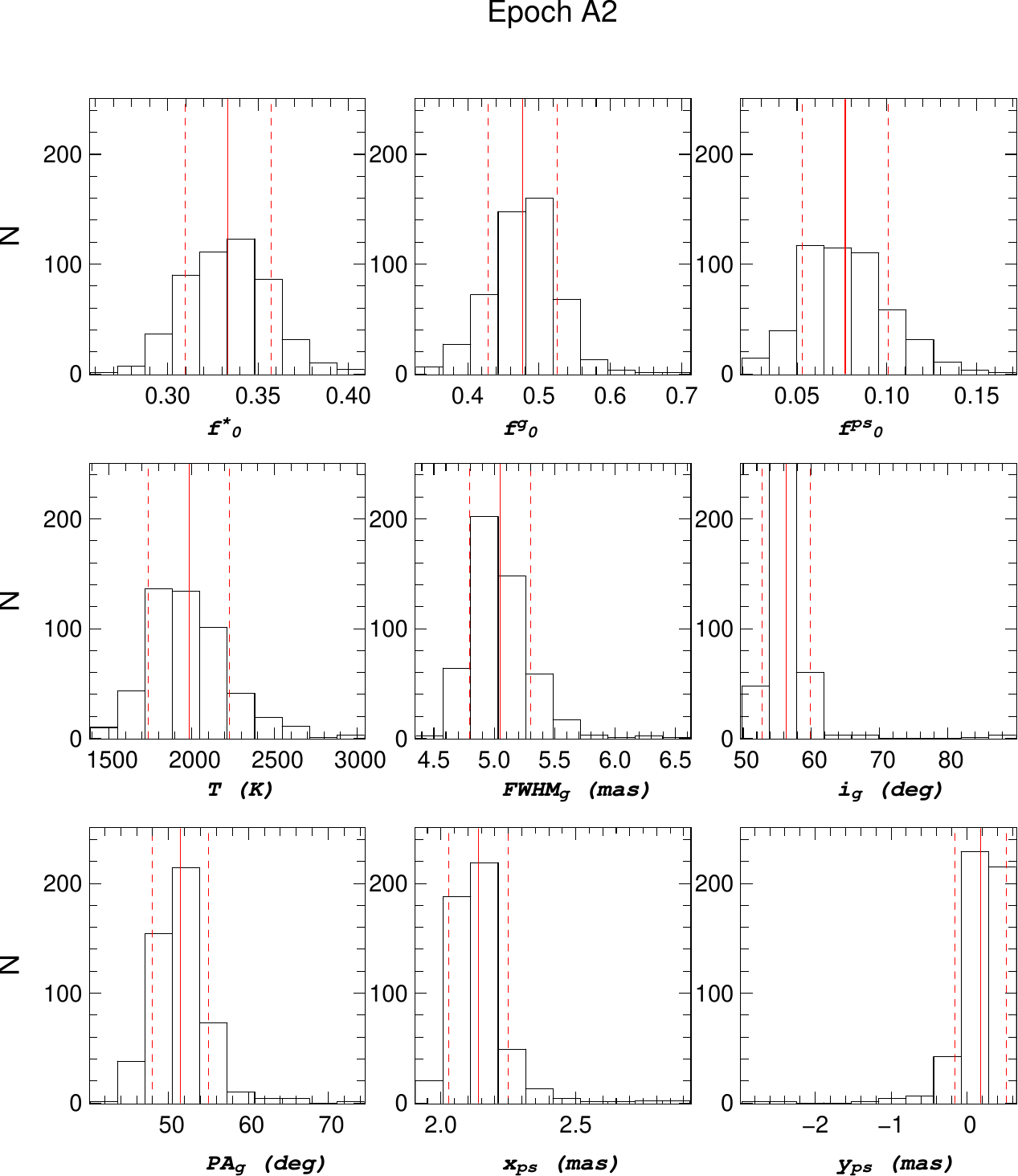} 
  \includegraphics[width=8cm]{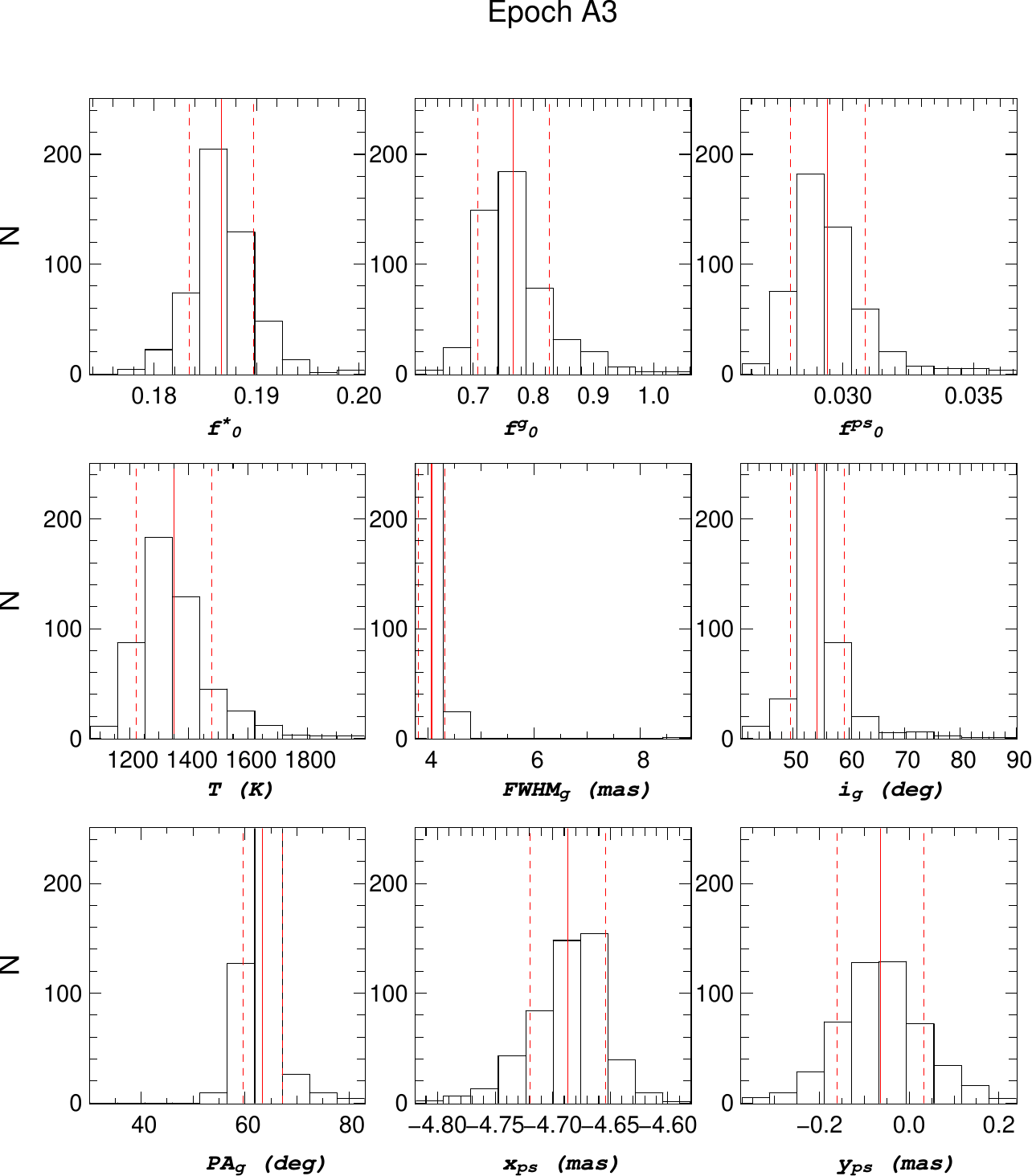} 
    \includegraphics[width=8cm]{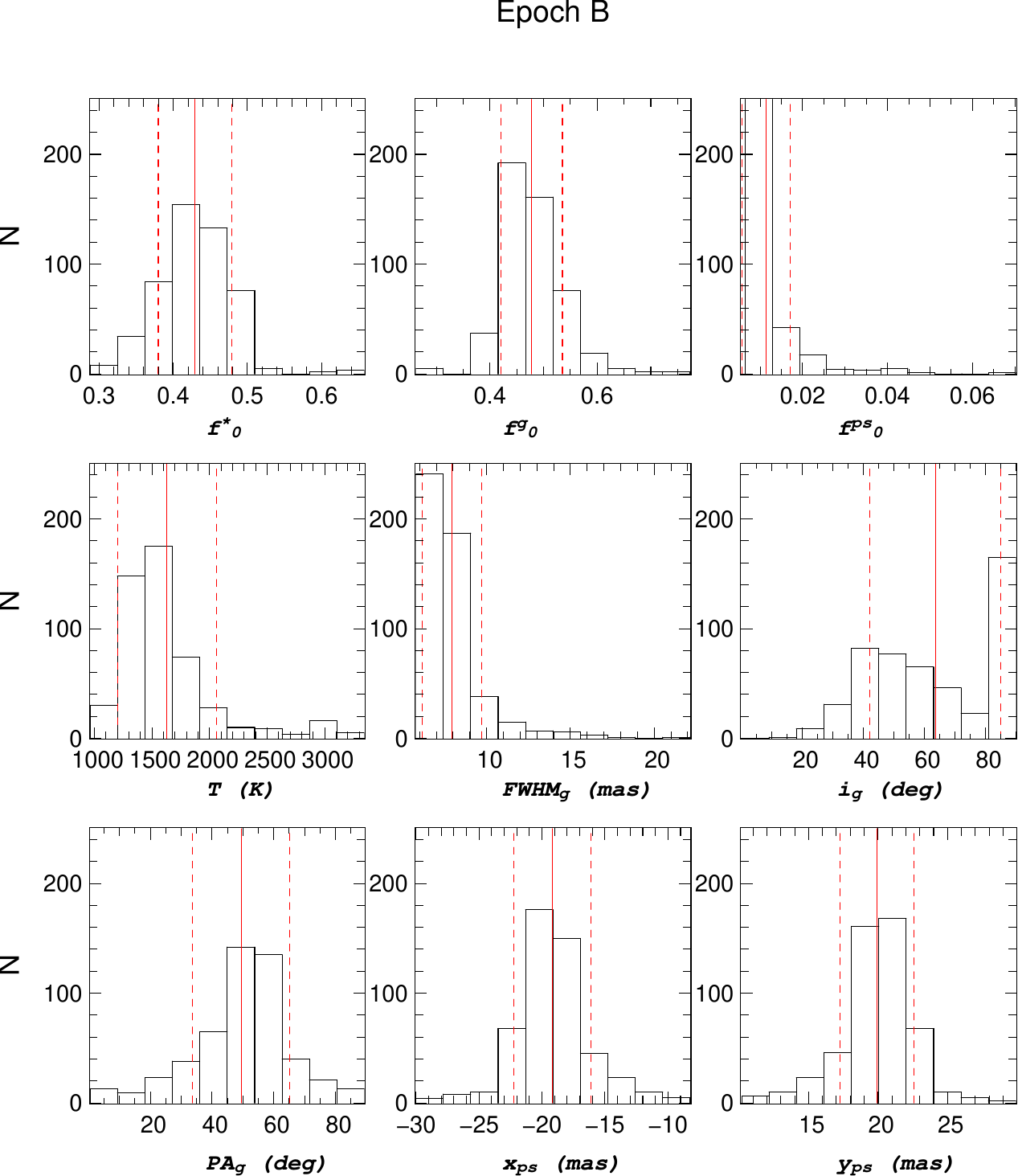} 
   \caption{Histograms of best-fit values for all parameters of the parametric fit for each epoch. The red solid line represents the mean values and the red dashed lines represent the 1 $\sigma$ errors.}
   \label{fig:paramerr}
\end{figure*}

\begin{figure*}[!t] 
   \centering
      \includegraphics[width=8cm]{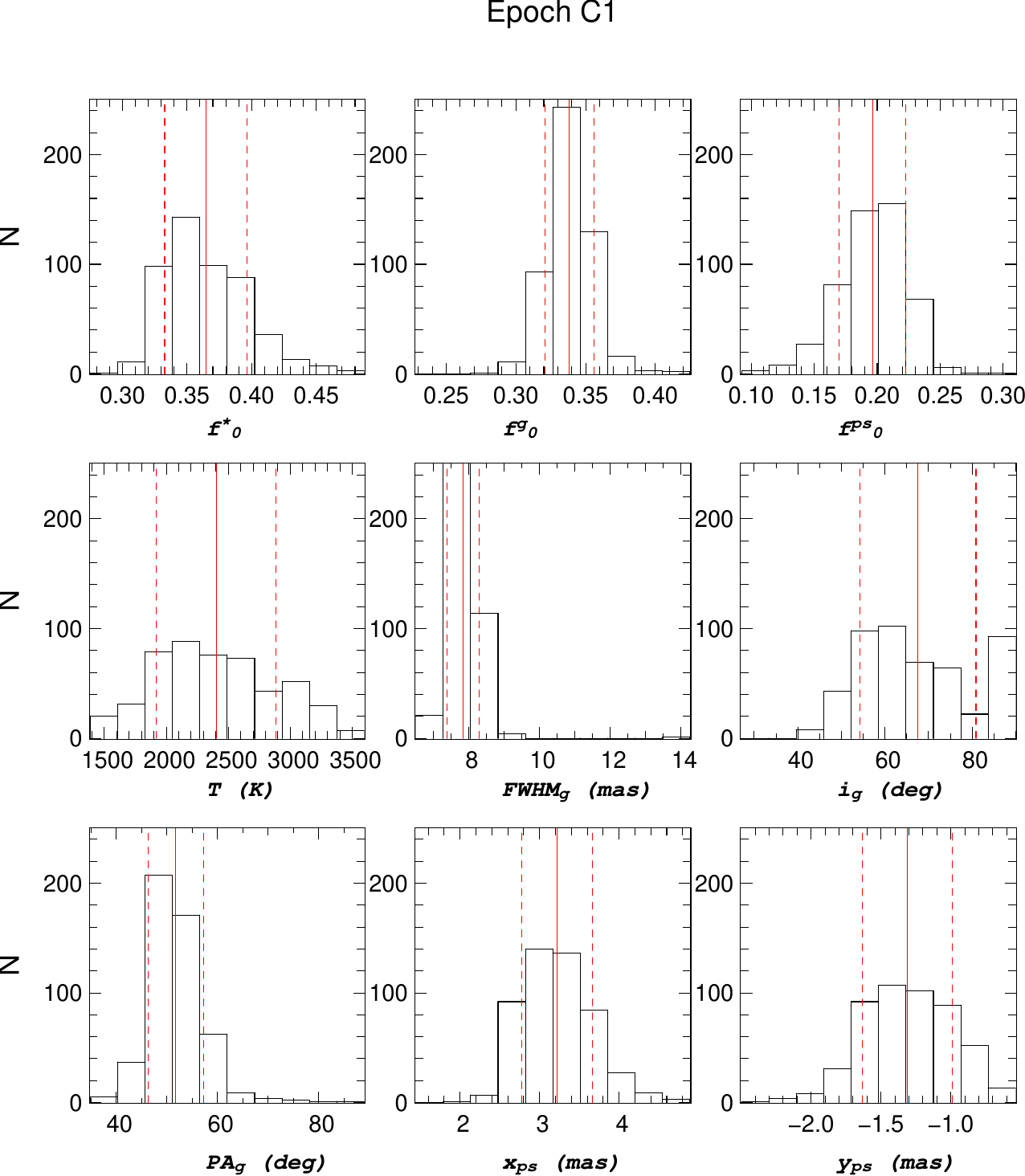} 
   \includegraphics[width=8cm]{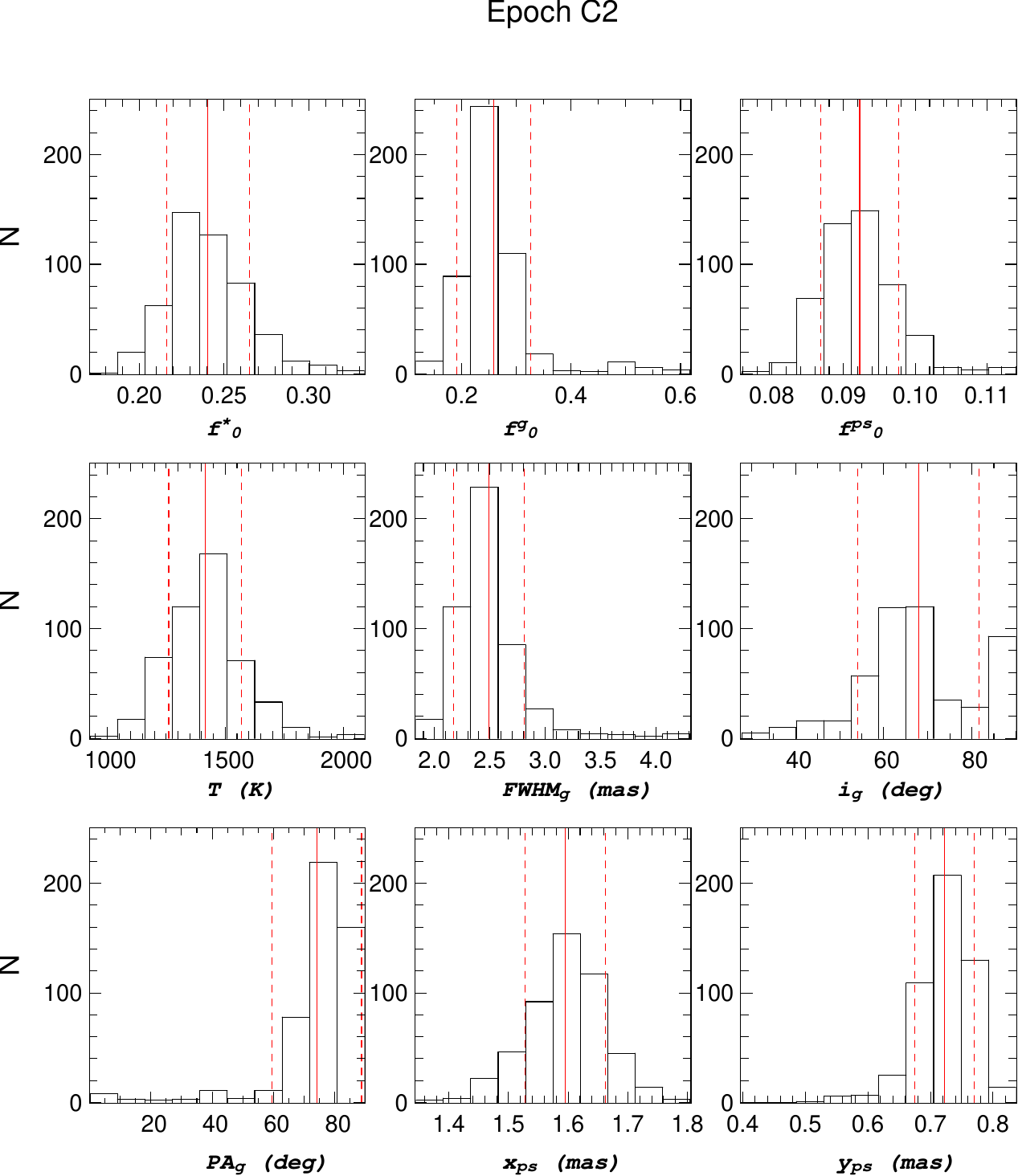} 
    \includegraphics[width=8cm]{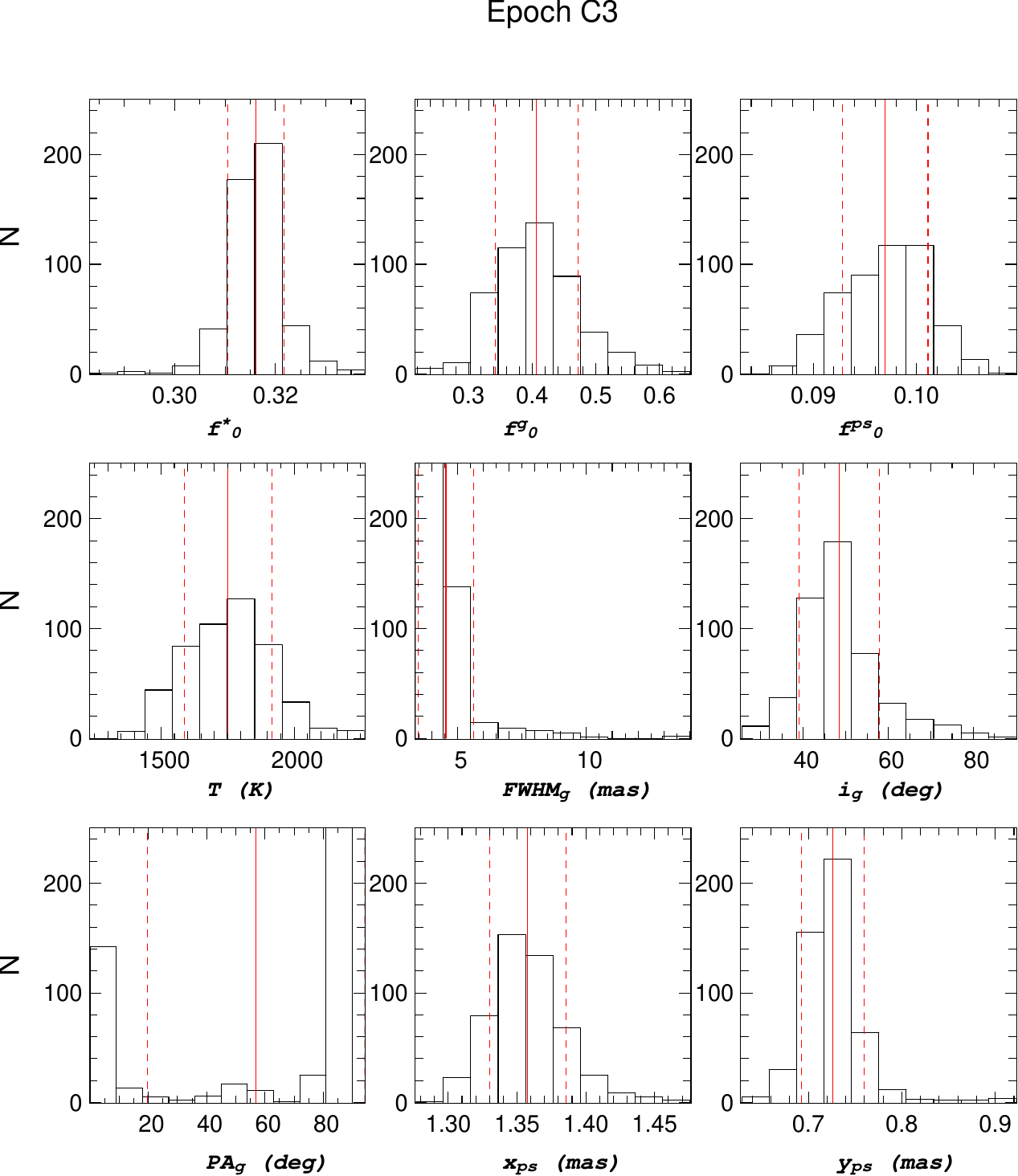} 
     \includegraphics[width=8cm]{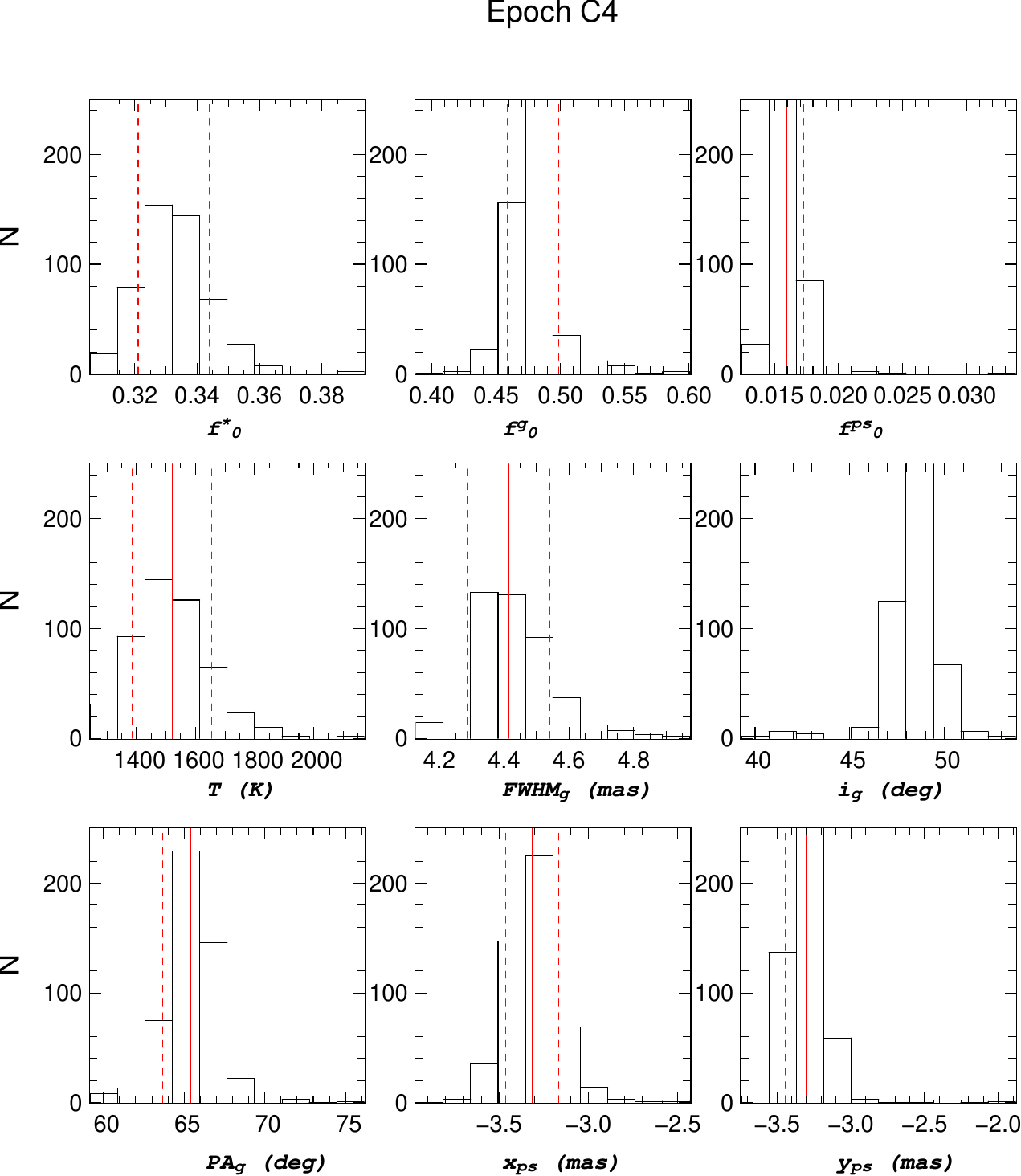} 
   \caption{Continuation of Fig.\,\ref{fig:paramerr}}
   \label{fig:paramerr2}
\end{figure*}

\section{Image reconstruction}
\label{app:rgl}
\begin{figure}[!t] 
   \centering
   \includegraphics[width=5.5cm]{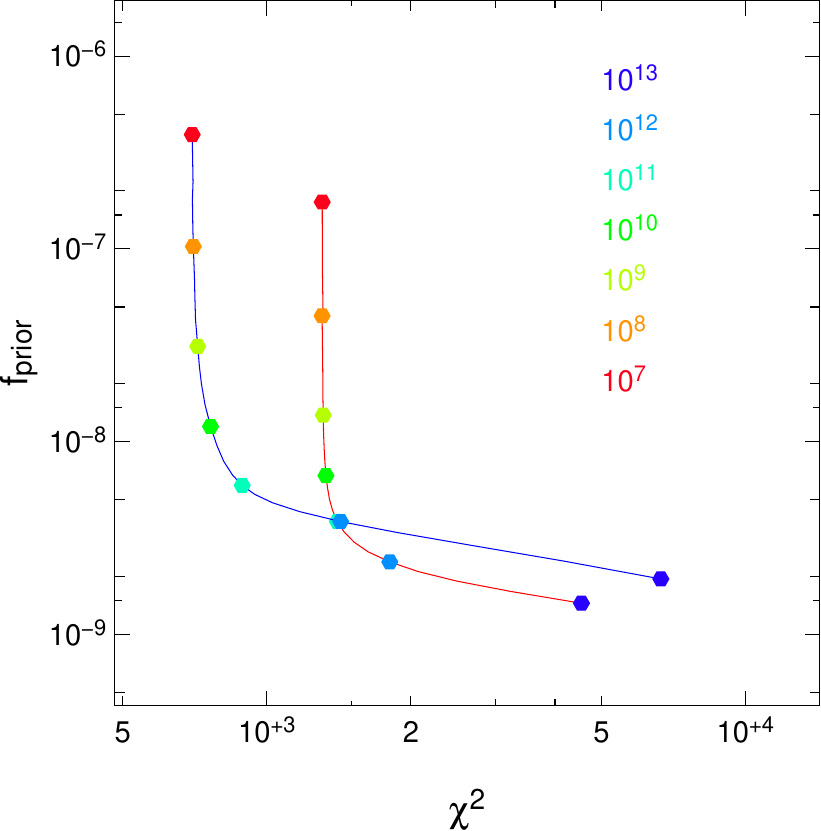} 
   \caption{Determination of the regularisation weight $\mu$ by the L-curve method. The reconstructions were performed on a grid of weights (logarithmic grid of 25 values from 10$^7$ to 10$^{13}$) for the two epochs (A in red and C in blue). For each reconstruction, the values of $\chi^2$ and $f_\mathrm{prior}$ were traced in this graph. The best value of the weight is that  in the shoulder of the curves. We have chosen 10$^{10}$ for both epochs.}
   \label{fig:Lcurve}
\end{figure}

The incomplete sampling of the Fourier plane forces us to bring \modif{in} additional information to reconstruct an image.
The regularisation plays the role of the missing information by promoting a morphology in the image (i.e. a smooth image).
In a Bayesian framework, these terms appear in the \textit{a posteriori} likelihood $P=\exp{-\frac{f}{2}}$ where $f = \chi^2 + \mu f_\mathrm{rgl}$, where $\chi^2$ is the mean square error, $f_\mathrm{rgl}$ is the regularisation function, and $\mu$ its weight,  \citep[see][for more details]{Thiebaut2008}, which we maximise to find the image.
Maximising the \textit{a posteriori} likelihood of the image amounts to minimising the function $f$.
We used the quadratic smoothness regularisation that promotes a smooth image and  is well suited for this object morphology \citep{Renard2011}.
We have set this weight to $10^{1\modif{0}}$ for both epochs, applying the L-curve method (see Fig.\,\ref{fig:Lcurve}).
It consists \modif{of} a grid of reconstructions with different weights of the regularisation.
The results are plotted on a $f_\mathrm{prior}$ versus $\chi^2$ plot. \modif{The resulting curve has} an L shape.
One asymptote is dominated by the data noise and the other by the regularisation.
The location of the bend indicates the optimal regularisation weight.

\begin{figure*}[!t] 
   \centering
   \includegraphics[width=10cm]{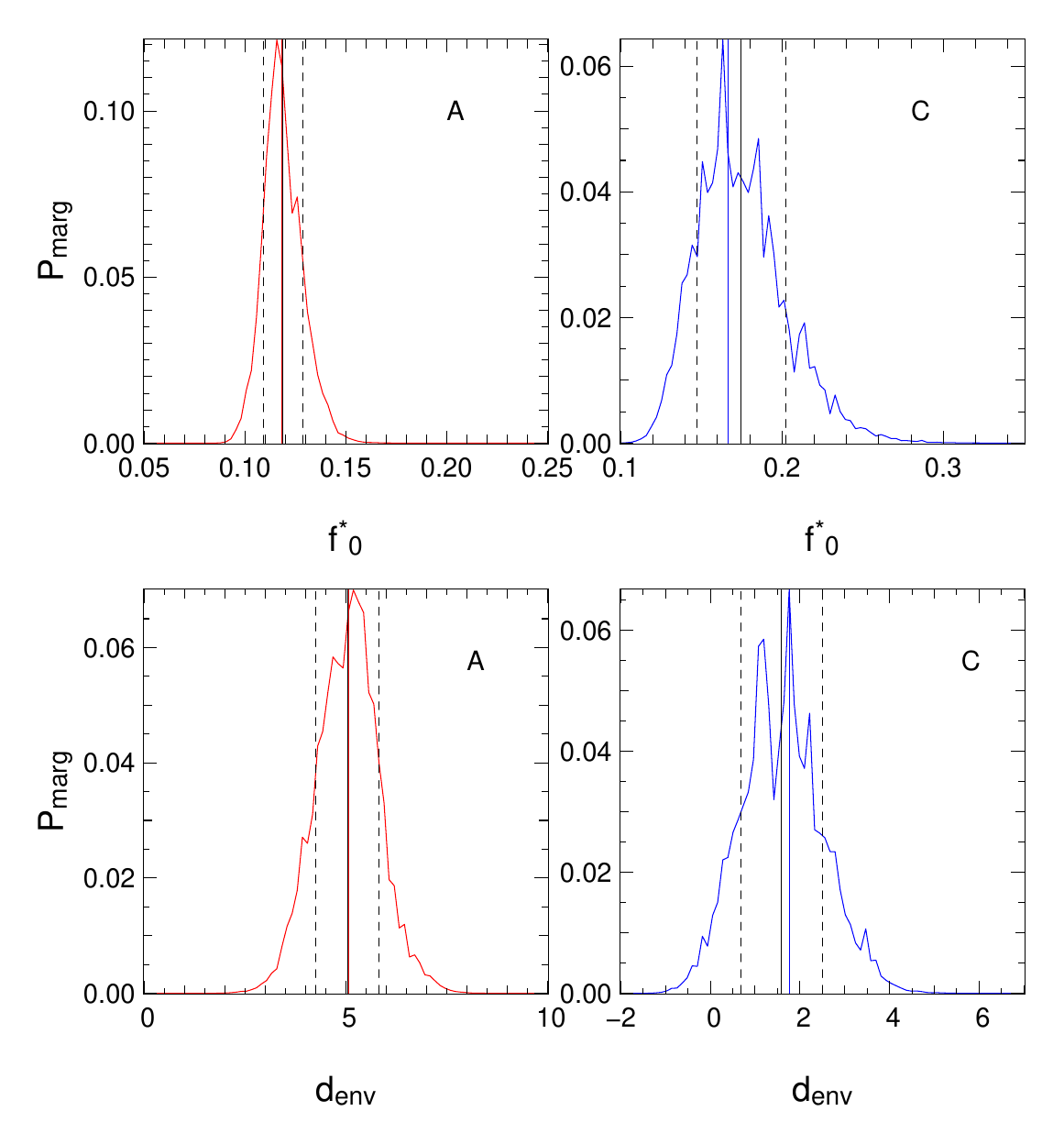} 
 \caption{Marginalised probabilities of the chromatic parameters \fso (top) and $d_\mathrm{env}$ (bottom) for epoch A (left) and C (right). The average value is indicated by solid vertical line and the standard deviation is indicated by dashed lines. The value chosen for the image reconstructions is represent by a vertical solid line, which is red \modif{for} epoch A and blue for epoch C.}
    \label{fig:marg}
\end{figure*}

\begin{figure*}[!t] 
   \centering
   \includegraphics[width=14cm]{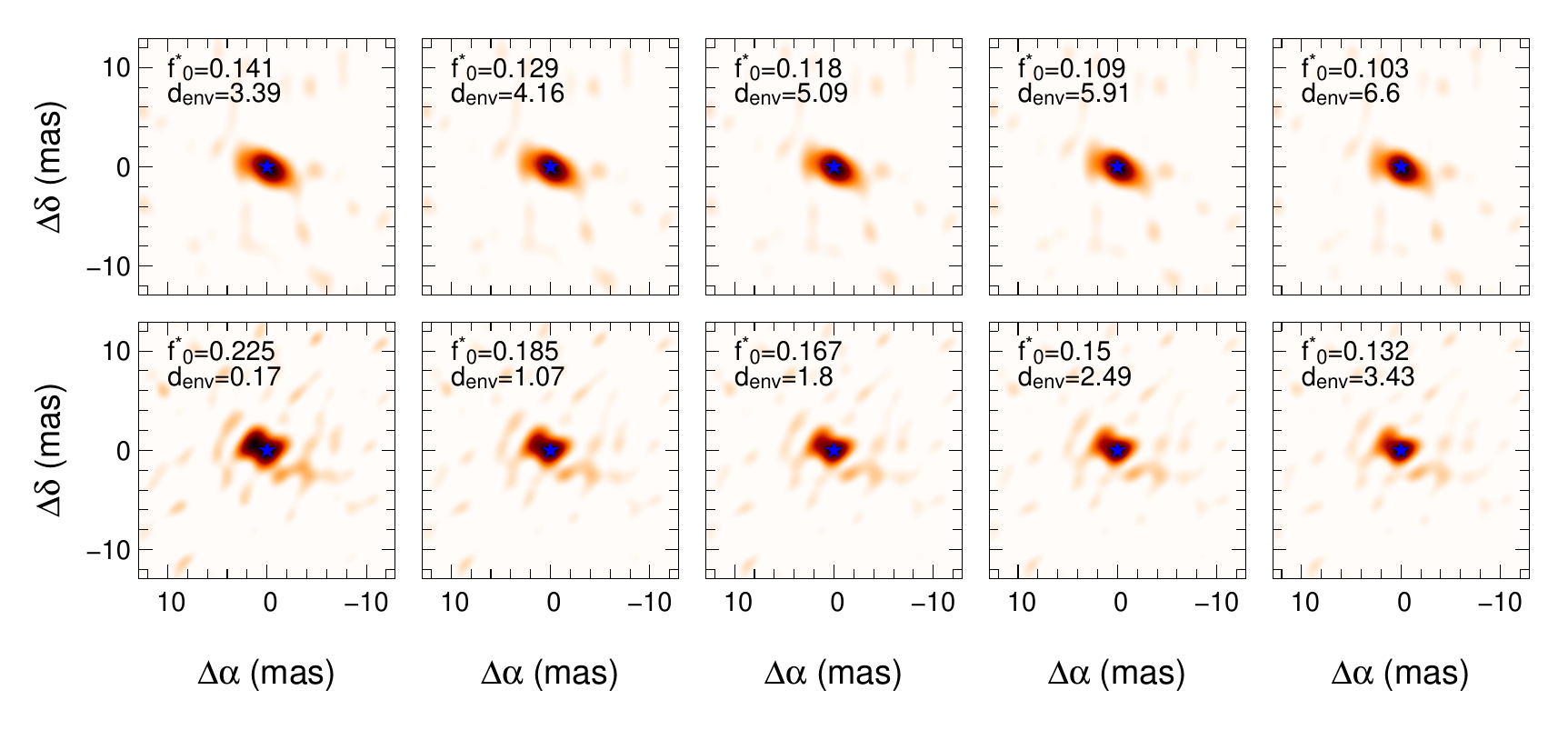} 
 \caption{Images for different values of the chromatic parameters in the 3 $\sigma$ contours of the Fig.\,\ref{fig:chrom}. The top line represent\modiff{s} the image reconstructions for epoch A and the bottom line the image reconstructions for epoch C. We can see no strong morphological changes for both epochs. For epoch C, however, we can see that the asymmetry in the north-east direction appear\modiff{s} with different brightness depending on the chromatic parameters.}
    \label{fig:mos}
\end{figure*}

\begin{figure*}[!t] 
   \centering
   \includegraphics[width=6cm]{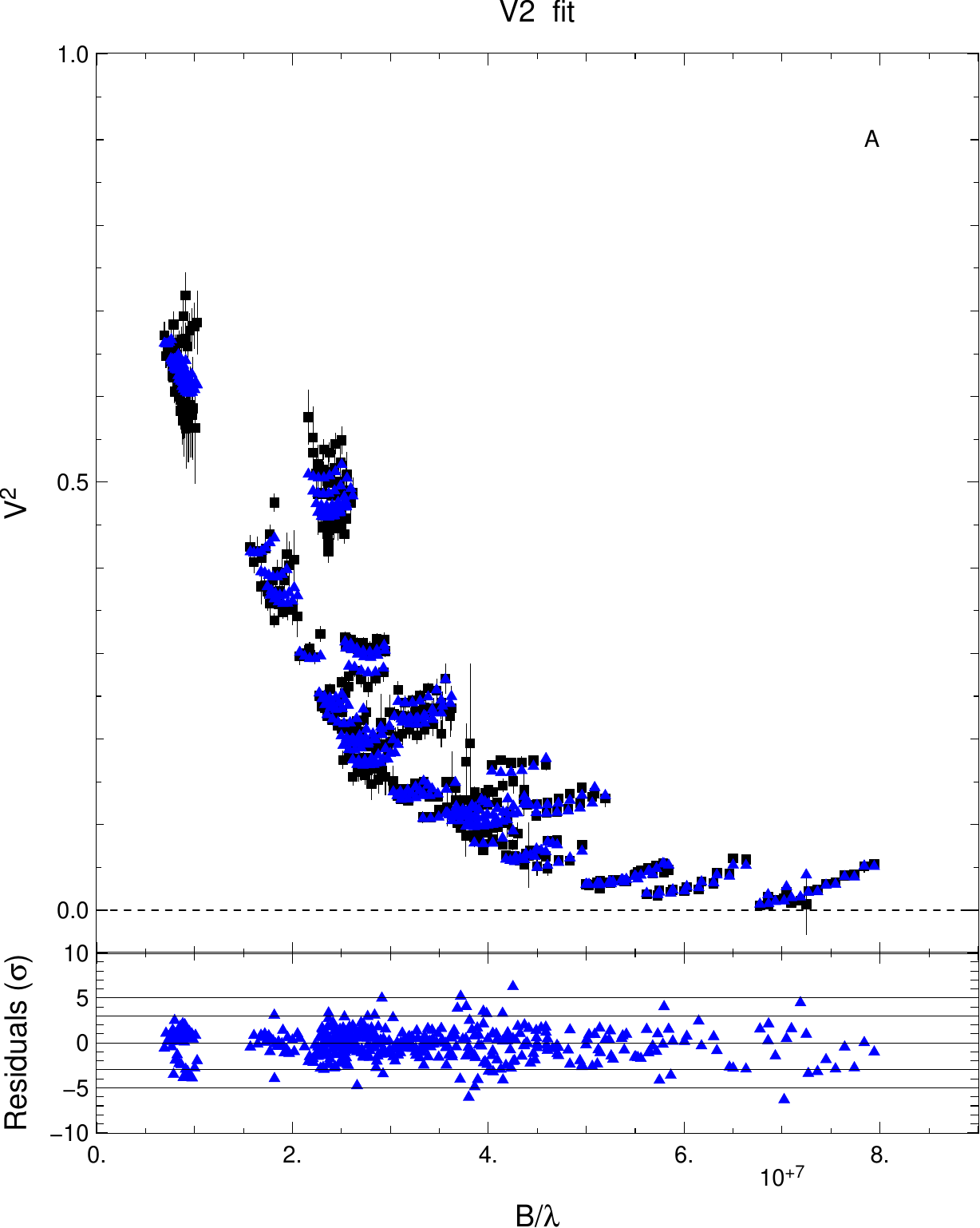} 
  \includegraphics[width=6cm]{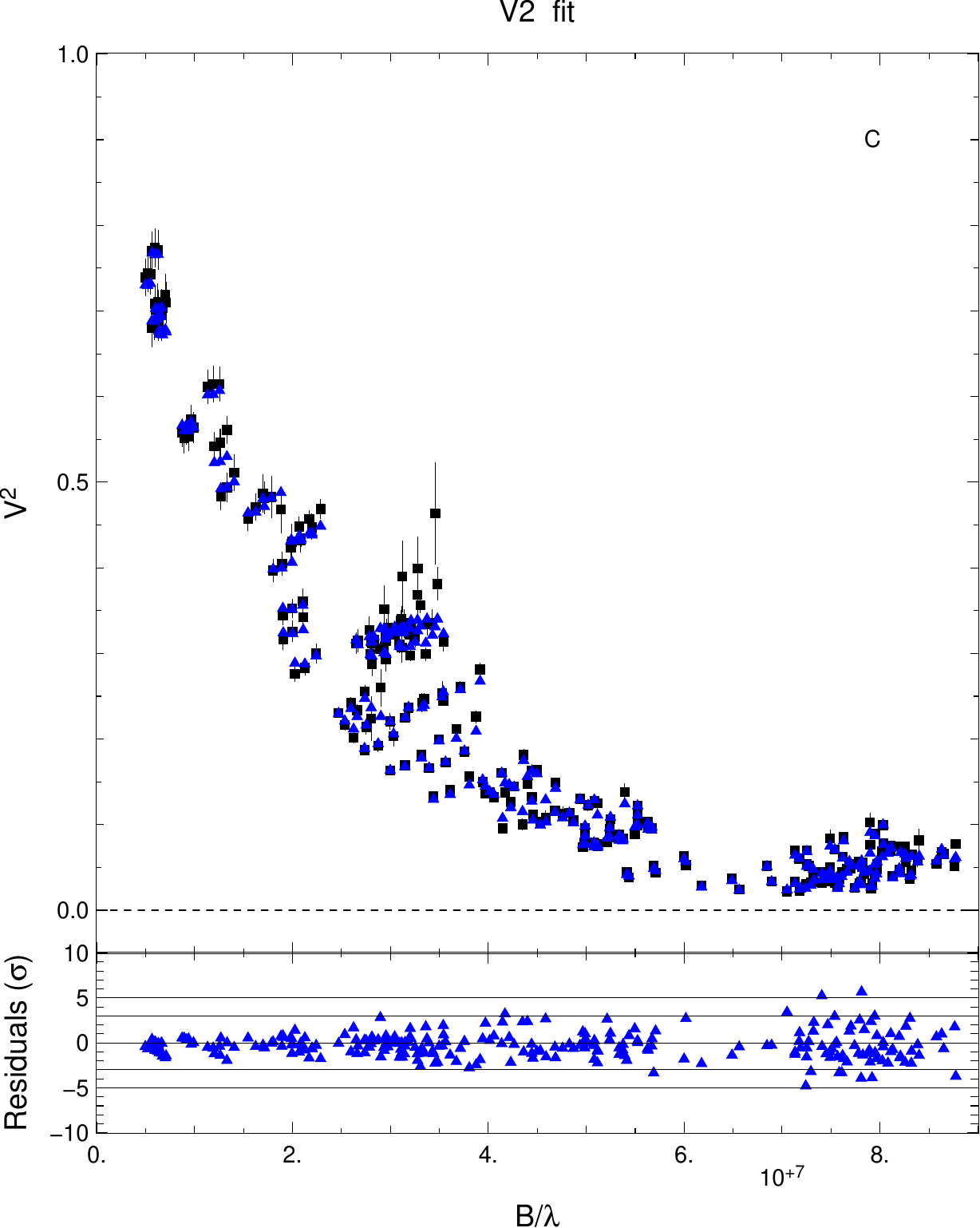} \\
   \includegraphics[width=6cm]{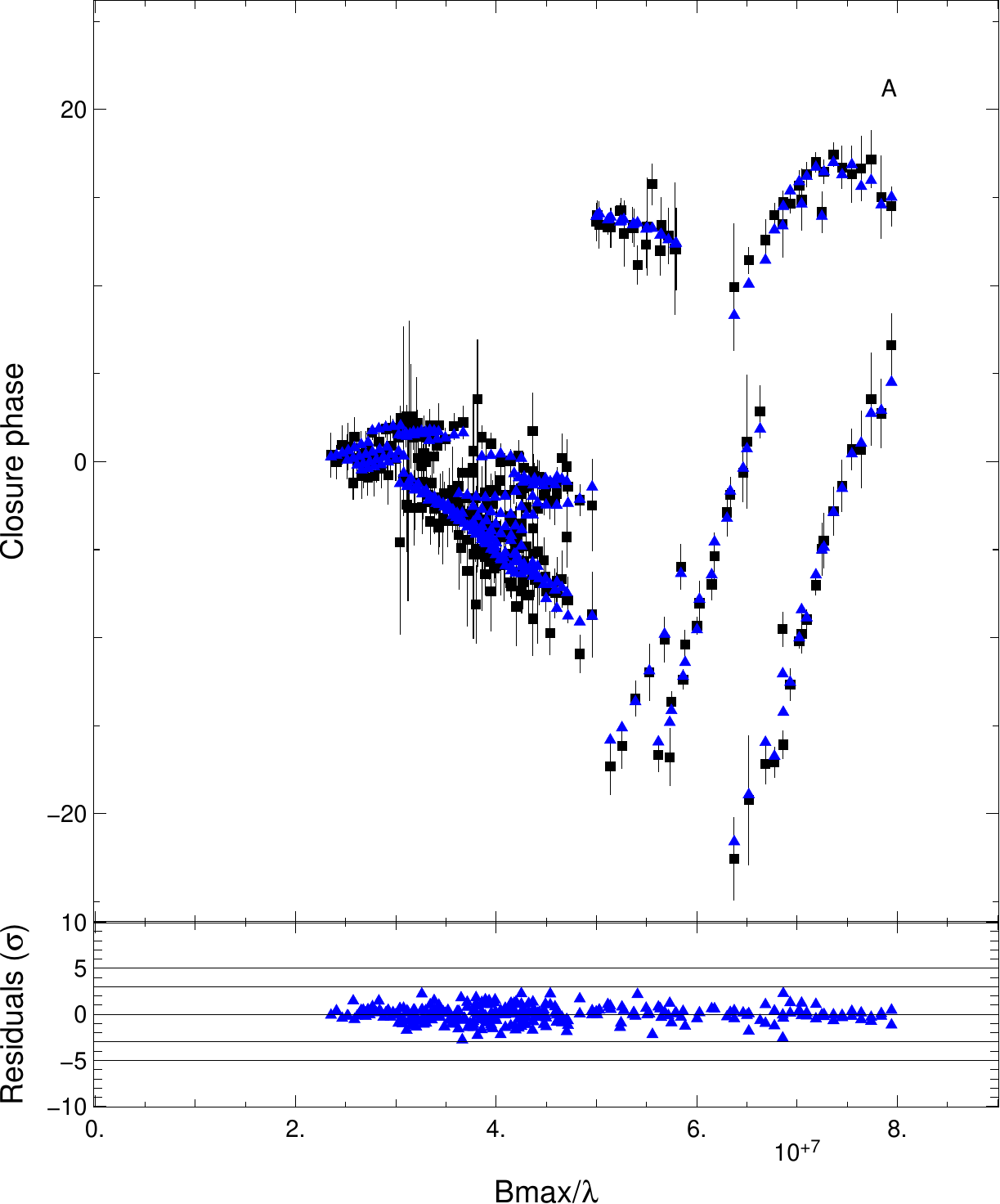} 
{\includegraphics[width=6cm]{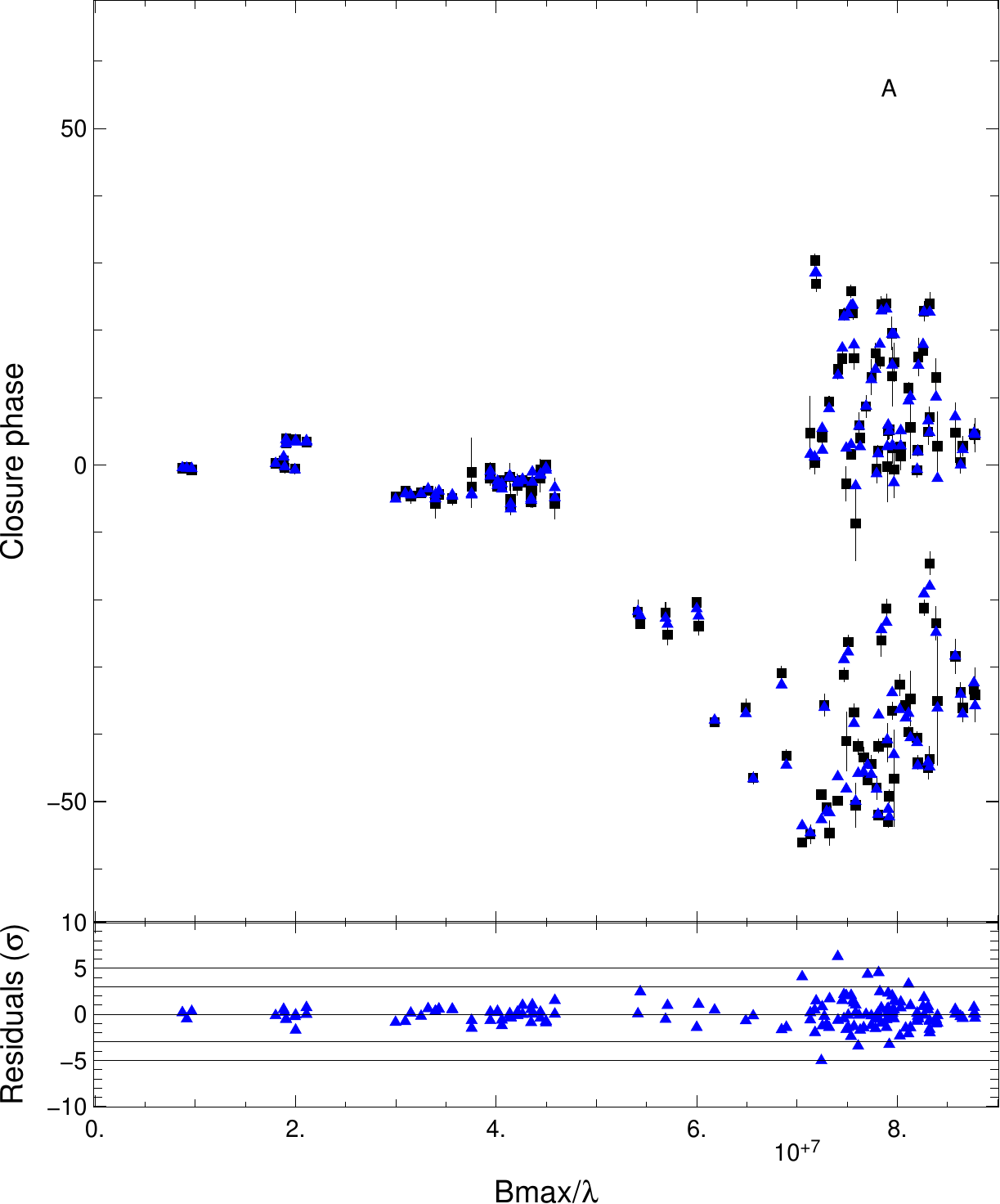}}
   \caption{Image reconstruction fit to the dataset on \VV (top) and on \modif{CP} (bottom) for epoch A (left) and C (right). The black squares represent the data and the blue triangles represent the image reconstruction fit. The residuals normalised to the standard deviation are plotted in the bottom of each graph. }
   \label{fig:qualimg}
\end{figure*}

\section{Trajectory of the \modif{asymmetry}}
\label{app:bin}

\begin{table}[t]
 \caption{Parameters of the best trajectory fit.} 
   \centering
   \begin{tabular}{@{} c|rr@{}} 
      $\chi^2_\mathrm{red}$  & 638  &\\
          \hline
      $a$ &5.2 &  (mas)\\
      $e$ & 0 & (fixed)\\
       $i$ &74.5& (deg)\\ 
      $\Omega$ & 70.9 &  (deg)\\
       $P$ & 88 & (days) \\
        $\Delta r$& 1.04 & (mas)
   \end{tabular}
   \label{tab:orbit}
   \tablefoot{$a$ is the semi-major axis, $e$ the eccentricity, $i$ the inclination, $\Omega$ the longitude of the ascending node (from north to east), $P$ the period, and $\Delta r$ the shift of the centre of the trajectory with respect to the star through the minor axis.}
\end{table}

\begin{figure}[!t] 
   \centering
     \includegraphics[width=6cm]{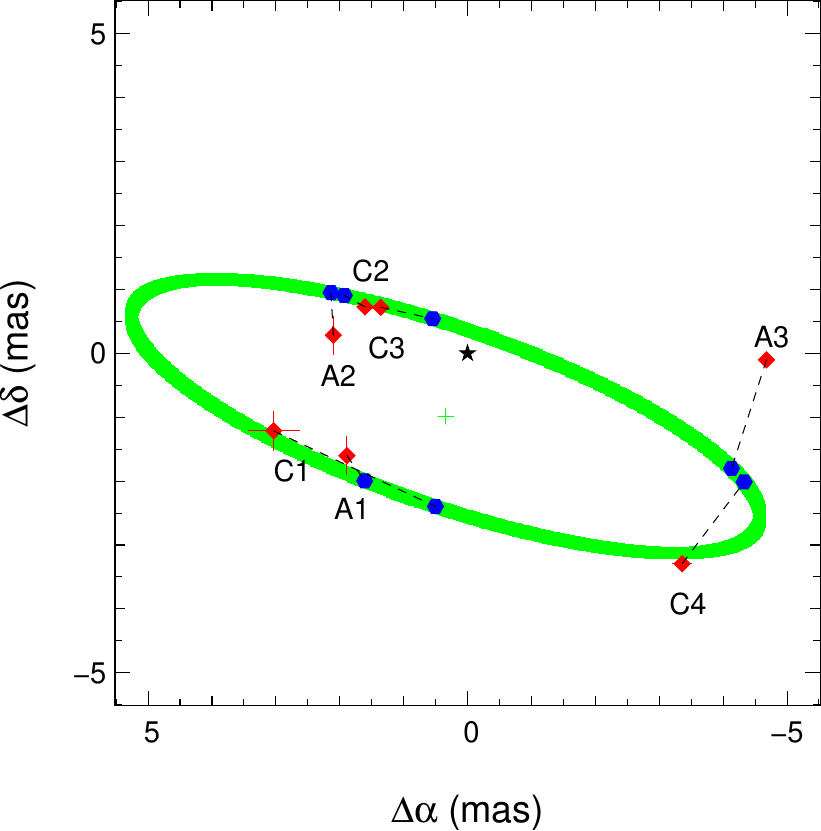} 
   \caption{Best-fit parameters of the circular trajectory. The red diamonds represent the positions of the \modif{asymmetry} from the best fits. The blue hexagons represent the positions of the \modif{asymmetry} for the best-fit trajectory. The black star represents the location of the central star. The green ellipse represents the fitted trajectory and the green cross shows its centre.}
\label{fig:KAP}
\end{figure}


We have fitted a circular trajectory to the position of the bright spot found in the parametric fit at all epochs but epoch B because the flux of the bright spot is not significant enough (see Fig.\,\ref{fig:KAP}).
This trajectory can be oriented on the plane of the sky.
We allow the centre of the trajectory to be shifted with respect to the central star along the minor axis ($\Delta r$).
Such a shift can be produced when observing an inclined disk with an \modif{asymmetry} orbiting the star above the equatorial plane.
The best values are presented Table\,\ref{tab:orbit}.

\modif{We obtained} a non-satisfactory reduced $\chi^2$ for the fit but \modif{it points out an} 88-day period \modif{with} a radius of 5.2 mas.
The orientation is in agreement with the values derived from the model shown in Sect.\,\ref{sec:paramfit} and from previous work \citep{BorgesII}.
There is a slight difference of 10 degrees on the inclination and the PA.
The centre of the trajectory shows a large shift with respect to the star such that the trajectory is passing close to the location of the star.

The image reconstruction epochs A and C are 56 and 63 days long.
This is of the order of the period of the bright spot.
As a consequence the reconstructed images are probably ``blurred".
The simple assumption of a displacement of a bright spot located in the disk is supported by the parametric analysis even if the models we are fitting are too simple to correctly reproduce the entire dataset, especially for the longest baselines.
\modif{Moreover, the orbit fit is imperfect owing to the poor temporal sampling and  its simplicity.
It cannot satisfactorily describe the asymmetry evolution.}

\bibliographystyle{aa}
\bibliography{biblio}

\end{document}